\documentclass[aps,prl,reprint,superscriptaddress,longbibliography]{revtex4-2}

\usepackage{times}

\usepackage{amsmath}
\usepackage{amsfonts}
\usepackage{amssymb}
\usepackage{mathtools}
\usepackage{bm}

\usepackage{graphicx}

\usepackage{booktabs}

\newcommand{\avg}[1]{\left\langle #1\right\rangle}
\newcommand{\HIC}{\mathrm{HIC}}
\newcommand{\NS}{\mathrm{NS}}
\newcommand{\CL}{\mathrm{CL}}
\newcommand{\ket}[1]{\left|#1\right\rangle}
\newcommand{\panel}[1]{\textbf{(#1)}}
\newcommand{\panelinclude}[3][]{%
\begingroup%
\setbox0=\hbox{\includegraphics[#1]{#2}}%
\raisebox{\dimexpr-\ht0\relax}[0pt][\dimexpr\ht0\relax]{\hbox{\copy0\hspace{-\wd0}%
\makebox[0pt][l]{\raisebox{\dimexpr\ht0-1.15em\relax}[0pt][0pt]{\hspace{0.18em}\footnotesize\panel{#3}}}%
\hspace{\wd0}}}%
\endgroup%
}

\usepackage[unicode=true,bookmarks=true,bookmarksnumbered=false,bookmarksopen=false,breaklinks=false,pdfborder={0 0 1},backref=false,colorlinks=true]{hyperref}
\hypersetup{linkcolor=blue,urlcolor=blue,citecolor=blue,pdfstartview={FitH},unicode=true}

\begin{document}

\title{Cluster-State Witnesses of Finite-Speed Hidden Influences}

\newcommand{\AffTsinghua}{Center for Quantum Information, IIIS, Tsinghua University, Beijing 100084, China}
\newcommand{\AffQizhi}{Shanghai Qi Zhi Institute, Shanghai 200232, China}
\newcommand{\AffHefei}{Hefei National Laboratory, Hefei 230088, China}
\newcommand{\AffCQT}{Centre for Quantum Technologies, National University of Singapore, 3 Science Drive 2, Singapore 117543, Singapore}
\newcommand{\AffDoP}{Department of Physics, National University of Singapore,  2 Science Drive 3, Singapore 117542, Singapore}

\author{Weikang Li}
\email{weikang\_li@tsinghua.edu.cn}
\affiliation{\AffTsinghua}

\author{Mengyao Hu}
\email{mengyao.hu@nus.edu.sg}
\affiliation{\AffCQT}

\author{Dong-Ling Deng}
\email{dldeng@tsinghua.edu.cn}
\affiliation{\AffTsinghua}
\affiliation{\AffQizhi}
\affiliation{\AffHefei}

\author{Valerio Scarani}
\email{physv@nus.edu.sg}
\affiliation{\AffCQT}
\affiliation{\AffDoP}

\begin{abstract}
Bell experiments rule out local common-cause explanations of quantum correlations, yet they do not exclude hidden influences that travel faster than light while still having a finite speed in a preferred frame.
Multipartite spacetime arrangements turn this possibility into a constraint: two late parties that are outside each other's hidden-influence cones must remain Bell-local once the earlier events are fixed. 
Here, we formulate this constraint as a projected-polytope separation problem for cluster-state correlations, using only marginal data containing at most one late party.
From linear cluster states, we construct a four-qubit witness with the bound \(S_4\le 6\) and quantum value \(4+2\sqrt2\), and a five-qubit witness with \(S_5\le 10\) and quantum value \(6+4\sqrt2\). 
We certify that the exposed faces are facets of the corresponding projected hidden-influence polytopes. 
These results identify linear-cluster graph states as certifiable and experimentally friendly resources for finite-speed hidden-influence tests.
\end{abstract}

\maketitle

\textit{Introduction.}---The Einstein-Podolsky-Rosen argument made locality and completeness a sharp problem for quantum theory~\cite{EPR1935}. Bell's theorem turned that problem into experimentally testable constraints on local common-cause explanations~\cite{Bell1964,CHSH1969,ClauserHorne1974,Brunner2014}. 
Experiments, from early tests to loophole-free implementations across platforms, have established Bell nonlocality under increasingly stringent conditions~\cite{FreedmanClauser1972,Aspect1982,Tittel1998,Weihs1998,Rowe2001,Scheidl2010,Handsteiner2017,Hensen2015,Giustina2015,Shalm2015,Rosenfeld2017,Storz2023}. These results rule out local common-cause explanations (under measurement independence, which we assume for the rest of the paper). Here we turn to the other mechanism for correlations, namely direct causation by a hidden influence propagating at speed $v$.

\begin{figure}[t]
\centering
\includegraphics[width=0.48\textwidth]{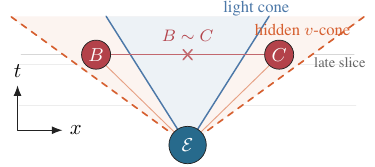}
\caption{\textbf{Finite-speed causality geometry.}
The blue solid lines denote the ordinary light cone, whereas the orange dashed lines denote the wider hidden \(v\)-cone associated with a finite influence speed \(v>c\) in the preferred frame.
The set \(\mathcal E\) denotes the early side, whose measurement event(s) may influence both late parties \(B\) and \(C\) through the hidden influence.
The late events are arranged outside each other's hidden \(v\)-cones, denoted by \(B\sim C\), so no hidden influence can propagate between them during a run.
Consequently, conditioned on every non-null input--output event on \(\mathcal E\), the \(BC\) distribution must be Bell-local.}
\label{fig:logic}
\end{figure}

The possibility that correlations can be explained with an unknown influence propagating at $v\leq c$ has been identified early on as the ``locality loophole''. It was closed in several experiments~\cite{Hensen2015,Giustina2015,Shalm2015,Rosenfeld2017,Li2018Test,Storz2023} by arranging spacelike separation between the choice of a measurement on one party and the production of the outcome of the other party (with operational definitions of these events). Next, one may consider hidden influences that propagate in a preferred frame at a superluminal but finite speed $c<v<\infty$~\cite{Eberhard1989}. The finiteness of $v$ introduces a $v$-cone, and a similar argument as for the locality loophole can then be made: if the model is correct, two events that are spacelike-separated for that cone (Fig.~\ref{fig:logic}) can only be correlated by common cause; if Bell inequalities are still violated, the $v$-influence model is falsified. Several dedicated bipartite experiments reported no disappearance of the Bell violation and the bound stands at $v\gtrsim 10^4c$~\cite{ScaraniTittel2000,SuarezScarani1997,Zbinden2001,Stefanov2002,Salart2008,Cocciaro2011,Yin2013,Cocciaro2018}. Putting such lower bounds is as far as bipartite scenarios can go.

Multipartite scenarios are essentially different when it comes to hidden influence models. 
It was first surmised~\cite{ScaraniGisin2002,ScaraniGisin2005}, then conclusively proved~\cite{Bancal2012,Barnea2013}, that in any hidden-influence model with $v<\infty$ (including therefore those with $v\leq c$) it becomes possible for the users to communicate faster than light. 
The mathematical formalization of the argument goes as follows. 
In a setting like that of Fig.~\ref{fig:logic}, the correlation between Bob and Charlie must be local, since the $v$-influence does not connect them, while their correlations $P_{\mathcal{E}B}$ and $P_{\mathcal{E}C}$ with the past $\mathcal{E}$ are quite arbitrary. Suppose now that \textit{all no-signaling} $P(o_{\mathcal{E}},b,c|s_{\mathcal{E}},y,z)\equiv P_{\mathcal{E}BC}$, whose marginals are $P(o_{\mathcal{E}},b|s_{\mathcal{E}},y)\equiv P_{\mathcal{E}B}$ and $P(o_{\mathcal{E}},c|s_{\mathcal{E}},z)\equiv P_{\mathcal{E}C}$, are such that $P(b,c|y,z)\equiv P_{BC}$ violates Bell. 
Then either the $v$-influence model fails to explain the violations and is therefore falsified, or $P_{\mathcal{E}BC}$ must be signaling and the parties could communicate faster than light (i.e.~the influence is not hidden)~\footnote{The protocol for actual signaling is obvious in the configurations of Refs.~\cite{ScaraniGisin2002,ScaraniGisin2005}. In Ref.~\cite{Bancal2012} there is a four-partite configuration with $\mathcal{E}=(A,D)$, in which signaling is enabled from $A$ to $D$. It has been argued that signaling correlations may be allowed if the spacetime configuration does not allow any users to exploit them~\cite{HorodeckiRamanathan2019,Weilenmann2025}.}. 
The specific spacetime configurations in which signaling can be activated may be hard to pinpoint, as one should first identify the relevant frame (frames, in some variants~\cite{Scarani2014}). 
At any rate, it is an inescapable feature of $v$-influence models: either the ``peaceful coexistence'' between Bell nonlocality and relativity is broken and faster-than-light communication can be activated; or the influence is instantaneous (i.e., $v=\infty$, instantaneous direct causation, as exemplified by Bohmian mechanics) or larger (retrocausation). 
These two last possibilities being unfalsifiable, we can say that any falsifiable direct causation in spacetime can be excluded, just as any form of common causation was excluded by Bell.

Despite the fact that this has been known for more than a decade, no experiment has reported even a proof-of-principle demonstration yet, because the requirements were too complex and tight. The hidden-influence witness of Bancal et al.~\cite{Bancal2012} could be tested with projective measurements on four qubits, but its maximal violation is small and the suitable states are rather complex. Barnea et al.~\cite{Barnea2013} found a three-partite example, but with an even smaller violation. In this work, we provide new four- and five-partite criteria to falsify all models with finite $v$, which have a much larger violation and can be tested with familiar states (namely, linear cluster states of qubits), paving the way for feasible experimental tests. 

\textit{Setting the stage.}--- Let \(\NS\) denote the set of full no-signaling distributions~\cite{PopescuRohrlich1994,Barrett2005}, and \(\CL(BC|\mathcal E)\) denote the set of full distributions for which
the \(BC\) distribution admits a Bell-local decomposition conditional on every non-null input-output event on the early side \(\mathcal E\).
The hidden-influence causality (HIC) set corresponding to the configuration of Fig.~\ref{fig:logic} is
\begin{equation}
\HIC_{\mathcal E}=\NS\cap\CL(BC|\mathcal E).
\label{eq:hic-set-main}
\end{equation}
It is strictly larger than the fully local Bell polytope because hidden influences from the early side to the late parties remain allowed, and is strictly smaller than the full no-signaling polytope because the blind pair must be local.

Finite-speed causality is therefore tested by designing a projected witness. The witness value is assembled from two marginal families: \(B\)-with-early and \(C\)-with-early. 
The measured functional is therefore no-\(BC\), meaning that every term containing both blind parties is omitted. 
Thus the relevant search object is the set
\begin{equation}
\mathcal P_{\mathcal E}^{\mathrm{no}\text{-}BC}
:=\underset{\mathrm{no}\text{-}BC}{\Pi}(\HIC_{\mathcal E})
=\underset{\mathrm{no}\text{-}BC}{\Pi}
\bigl(\NS\cap\CL(BC|\mathcal E)\bigr).
\label{eq:projected-hic-main}
\end{equation}
In the inequality \(S\le 7\) of Ref.~\cite{Bancal2012}, the reported quantum violation reaches only \(7.2014\), a gap of \(2.9\%\) when normalized by the bound; equivalently, the ideal white-noise threshold is \(0.9720\). Even the optimized value \(7.3481\) leaves a threshold \(0.9526\).
Considering finite statistics, state preparation errors, calibration drift, and the fact that the expression contains 23 nonuniform correlator terms, the witness is not a compact experimental target for near-term devices.

\begin{figure*}[t]
\centering
\begin{minipage}[t]{0.495\textwidth}
\vspace{0pt}
\centering
\begin{tabular}{@{}c@{}c@{}}
\panelinclude[width=0.468\linewidth]{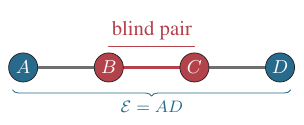}{a} &
\panelinclude[width=0.532\linewidth]{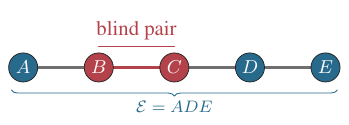}{b}\\
\multicolumn{2}{@{}l@{}}{\panelinclude[width=\linewidth]{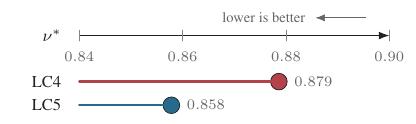}{c}}
\end{tabular}
\end{minipage}\hfill%
\begin{minipage}[t]{0.495\textwidth}
\vspace{2.7mm}
\centering
\panelinclude[width=\linewidth,trim=0 0 0 -2.7mm]{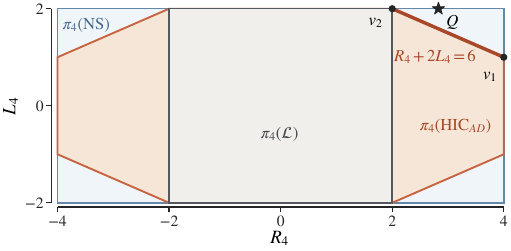}{d}
\end{minipage}
\caption{\textbf{Linear-cluster witnesses, visibility, and illustrative HIC projection.}
\panel{a} LC4 uses the four-qubit chain with early side \(\mathcal E=AD\) and blind late pair \(B,C\).
\panel{b} LC5 appends \(E\), giving \(\mathcal E=ADE\) with the same blind pair. The \(BC\) link in each panel denotes a graph-state edge, not a causal connection; no witness term contains both \(B\) and \(C\).
\panel{c} Ideal white-noise visibility, given by the ratio of the HIC bound to the quantum value \(S^Q\), with lower values indicating greater noise tolerance.
\panel{d} LC4 projection \(\pi_4(P):=(R_4(P),L_4(P))\). The fully Bell-local set \(\mathcal L\), hidden-influence set \(\HIC_{AD}\), and no-signaling set \(\NS\) project to the inner square, intermediate octagon, and outer rectangle, respectively. The highlighted HIC facet \(R_4+2L_4=6\) joins the attainable projected vertices \(v_2=(2,2)\) and \(v_1=(4,1)\); the other three slanted facets follow from independent global output flips at \(B\) and \(C\). The cluster-state point \(Q=(2\sqrt2,2)\) lies on the boundary of \(\pi_4(\NS)\) but outside \(\pi_4(\HIC_{AD})\). Panel \panel{d} shows this exact two-dimensional block projection, not the full \(44\)-dimensional no-\(BC\) HIC projection.}
\label{fig:witnesses}
\end{figure*}

Here, we address this problem by constructing witnesses from linear cluster states,
motivated by stabilizer and graph-state methods~\cite{CoiteuxRoy2025,Raussendorf2001,Hein2004,Guhne2005,Hein2006} and the multipartite nonlocality  viewpoint~\cite{Mermin1990,Svetlichny1987,WernerWolf2001,Sliwa2003,Coretti2011}.
We choose sparse no-\(BC\) stabilizer directions and certify their hidden-influence support bounds against the projected polytope using integer Farkas certificates~\cite{Schrijver1986}; exact rational-rank and relative-interior checks then certify facet tightness~\cite{Ziegler1995,Pironio2005}. 
This yields two compact witnesses: the four-qubit witness LC4 uses six no-\(BC\) correlators and no four-body terms, while the five-qubit witness LC5 trades higher-weight correlators for a lower ideal white-noise threshold (Fig.~\ref{fig:witnesses}(a-c)). 
Both give substantially larger margins than previous finite-speed hidden-influence witnesses. They also involve few correlators, and can be maximally violated by simple two-setting measurements in the \(X\)--\(Z\) plane on linear-cluster states. All these features make them more experimentally tractable targets.

\textit{Projected hidden-influence polytope~\cite{Barnea2013}.}---To characterize \(\mathcal P_{\mathcal E}^{\mathrm{no}\text{-}BC}\), we begin with the full conditional-local distribution before applying the no-\(BC\) projection. Let \(s_{\mathcal E},o_{\mathcal E}\) denote the binary input and output strings of the early side \(\mathcal E\), and \(y,z\) (\(b,c\)) the inputs (outputs) of \(B,C\). In a \(B\sim C\) configuration, conditional locality requires
\begin{equation}
P(o_{\mathcal E},b,c|s_{\mathcal E},y,z)
=
\sum_{\lambda}
\tau^{s_{\mathcal E}}_{o_{\mathcal E},\lambda}
P(b|y,\lambda)P(c|z,\lambda).
\label{eq:conditional-locality}
\end{equation}
Here \(\lambda\) is a shared hidden variable and \(\tau^{s_{\mathcal E}}_{o_{\mathcal E},\lambda}\ge0\) are subnormalized weights satisfying \(\sum_{\lambda}\tau^{s_{\mathcal E}}_{o_{\mathcal E},\lambda}=P(o_{\mathcal E}|s_{\mathcal E})\) by no-signaling. Only the \(B,C\) responses factorize; the full distribution remains subject to no-signaling.
The deterministic response pairs specify only the \(B,C\) responses and do not by themselves define feasible distributions. 
Since \(\mathcal P_{\mathcal E}^{\mathrm{no}\text{-}BC}\) is obtained by projecting the distributions allowed by the constrained weights, enumerating those pairs alone does not determine its HIC bound.

Thus, we optimize each chosen no-\(BC\) functional directly over the constrained weights and certify its HIC bound. All retained terms can be evaluated from marginals accessible in the prescribed timing protocol. For \(\mathcal E=AD\) in LC4 and \(\mathcal E=ADE\) in LC5, we write the deterministic \(B,C\) responses as \(\beta,\gamma:\{0,1\}\to\{0,1\}\), and let \(t^{s_{\mathcal E}}_{o_{\mathcal E},\beta,\gamma}\ge0\) denote the subnormalized weight of the early output \(o_{\mathcal E}\) and response pair \((\beta,\gamma)\), given \(s_{\mathcal E}\). Equation~(\ref{eq:conditional-locality}) then becomes
\begin{equation*}
P(o_{\mathcal E},b,c|s_{\mathcal E},y,z)
=\sum_{\beta,\gamma}t^{s_{\mathcal E}}_{o_{\mathcal E},\beta,\gamma}
\mathbf{1}[b=\beta(y)]\mathbf{1}[c=\gamma(z)] .
\end{equation*}
Let \(t\) collect these weights. Normalization and no-signaling impose
\(\mathsf Nt=r\), with \(t\ge0\). Here \(\mathsf N\) is the constraint matrix, and \(r\) has entries \(1\) for normalization constraints and \(0\) for no-signaling constraints.
Each functional is linear in \(t\) and can therefore be written as \(S=\alpha^Tt\), where \(\alpha\) is its coefficient vector. Any vector \(\eta\) satisfying
\begin{equation}
\mathsf N^T\eta\ge\alpha,\qquad r^T\eta=\mathcal B
\label{eq:dual-main}
\end{equation}
certifies \(S\le\mathcal B\)~\cite{Schrijver1986}, since \(t\ge0\) implies \(S\le\eta^T\mathsf Nt=\eta^Tr=\mathcal B\). The full derivation is given in Secs.~III-V of the Supplemental Material, thereby establishing the exact bounds.

\textit{Construction from linear cluster states.}--- 
Our use of linear cluster states differs from the usual graph-state Bell program, such as testing ordinary local realism~\cite{Guhne2005,Toth2006}, certifying entanglement through stabilizer witnesses~\cite{TothGuhne2005PRL,TothGuhne2005PRA}, or supporting self-testing statements~\cite{McKague2014,Baccari2020}.
Stabilizers provide deterministic Pauli correlations, while a \(45^\circ\) rotation of the \(B\) measurement basis converts selected stabilizers into the \(1/\sqrt{2}\) correlations of a CHSH block.
Coiteux-Roy \textit{et al.} used a closely related stabilizer-CHSH construction for a four-qubit cluster state under a different model, which provides a technical inspiration for this work~\cite{CoiteuxRoy2025}.
In our setting, the no-\(BC\) constraint forbids joint \(BC\) correlators, but not a \(BC\) edge in the prepared graph.  
The witness must be assembled from \(B\)-with-early and \(C\)-with-early marginals.  
We therefore combine a \(B\)-side CHSH block with \(C\)-side stabilizer locks and determine their relative weights from the exact HIC support.

\textit{Result 1: LC4.}---Consider the four-qubit linear cluster state
\begin{equation*}
\ket{LC_4}=CZ_{AB}CZ_{BC}CZ_{CD}\ket{+}^{\otimes 4},
\end{equation*}
as illustrated in Fig.~\ref{fig:witnesses}(a). We choose
\begin{gather*}
A_0=X_A,\quad A_1=Z_A,\quad
C_0=Z_C,\quad C_1=X_C,\\
D_0=X_D,\quad D_1=Z_D .
\end{gather*}
Four no-\(BC\) stabilizer products enter the construction: \(X_AZ_B\), \(Z_AX_BX_D\), \(Z_CX_D\), and \(X_AX_CZ_D\), each with expectation value \(+1\) on \(\ket{LC_4}\) (see Supplemental Material, Sec.~VI). The first two can be written as \(U_0Z_B\) and \(U_1X_B\), where \(U_0:=A_0\) and \(U_1:=A_1D_0\) are dichotomic functions of the early-side outcomes and serve as the two effective CHSH variables. Choosing
\begin{equation*}
\sqrt2B_0=Z_B+X_B,\qquad
\sqrt2B_1=Z_B-X_B
\end{equation*}
gives \(B_0+B_1=\sqrt2 Z_B\) and \(B_0-B_1=\sqrt2 X_B\). The first pair therefore forms the distributed CHSH block \(R_4\), while the remaining two stabilizers define the lock block \(L_4\):
\begin{align*}
R_4&:=\avg{A_0(B_0+B_1)}
      +\avg{A_1D_0(B_0-B_1)},\\
L_4&:=\avg{C_0D_0}+\avg{A_0C_1D_1}.
\end{align*}
The cluster state realizes \((R_4,L_4)=(2\sqrt2,2)\). For dichotomic outcomes, \(L_4=2\) enforces \(C_0=D_0\) and \(C_1=A_0D_1\) in their respective measurement contexts.

The CHSH block alone cannot witness HIC because the early side \(AD\) may influence \(B\), allowing \(R_4\) to reach its algebraic maximum \(4\). To optimize the block weights, we project \(\HIC_{AD}:=\NS\cap\CL(BC|AD)\) onto the \((R_4,L_4)\) plane. The projected set contains two attainable boundary points,
\(v_1=(4,1)\) and \(v_2=(2,2)\). Here \(v_i\) denotes the image of a full HIC distribution \(P_i\in\HIC_{AD}\) under the projection
\(P\mapsto(R_4(P),L_4(P))\); explicitly,
\[
(R_4(P_1),L_4(P_1))=(4,1),\quad
(R_4(P_2),L_4(P_2))=(2,2).
\]
The corresponding full distributions are constructed in Supplemental Material, Sec.~II~B.
\(v_1\) and \(v_2\) capture the trade-off between maximizing the CHSH block \(R_4\) and maximizing the stabilizer-lock block \(L_4\).
Their difference \(v_2-v_1=(-2,1)\) has the primitive outward normal \((1,2)\), giving the candidate supporting inequality \(R_4+2L_4\le6\), with \(v_1\) and \(v_2\) showing that it is tight; see Fig.~\ref{fig:witnesses}(d). Together with the algebraic bounds \(R_4\le4\) and \(L_4\le2\), these constraints imply
\begin{equation*}
\max_{P\in\HIC_{AD}}\bigl(pR_4+\ell L_4\bigr)
=\max\{4p+\ell,\,2p+2\ell\},\quad p,\ell\ge0.
\end{equation*}
The two branches coincide at \(\ell=2p\). For the cluster state, the ratio of \(2\sqrt2\,p+2\ell\) to the HIC support increases up to this intersection and decreases thereafter. Thus \((p,\ell)\propto(1,2)\) gives the largest cluster-state violation relative to the HIC bound, yielding the optimal choice \(R_4+2L_4\le6\) (Fig.~\ref{fig:witnesses}(d)).

Expanding this inequality, we have
\begin{equation}
\begin{aligned}
S_4={}&
\avg{A_0B_0}+\avg{A_0B_1}
+\avg{A_1B_0D_0}-\avg{A_1B_1D_0}\\
&+2\avg{C_0D_0}+2\avg{A_0C_1D_1}
\le 6 .
\end{aligned}
\label{eq:S4-main}
\end{equation}
The supports are $AB$, $ABD$, $CD$, and $ACD$, so the expression is strictly no-\(BC\).  It uses six two- and three-body correlators and a maximum coefficient of $2$. Its quantum value and white-noise visibility are 
\begin{align*}
S_4^Q&=4+2\sqrt2\approx6.8284,\\
\nu_4^*&=\frac{6}{4+2\sqrt2}=\frac{6}{S_4^Q}\approx0.8787,
\end{align*}
because $S_4=0$ on white noise. $\HIC_{AD}$ and fully local polytopes are distinct, as illustrated in Fig.~\ref{fig:witnesses}(d). Interestingly, $S_4=6$ is also the full local bound (achieved e.g.~by $A_0=+1$, any $A_1$, $B_0=B_1=+1$, $C_0=D_0$ and $C_1=D_1$). In other words, $S_4$ is insensitive to nonlocal correlations other than those involving both B and C. Over the full no-signaling set, the expression reaches its algebraic maximum $S_4=8$. 

\begin{table}[t]
\caption{\textbf{Certified hidden-influence witnesses.}  Dimensions are computed after projection to no-\(BC\) correlator coordinates.  The tightness column reports the projected affine dimension and the active-face dimension.}
\label{tab:main-numbers}
\begin{ruledtabular}
\begin{tabular}{lcccc}
model & HIC bound & quantum value & $\nu^*$ & tightness\\
\hline
LC4 & $6$ & $4+2\sqrt2$ & $0.8787$ & $(44, 43)$ \\
LC5 & $10$ & $6+4\sqrt2$ & $0.8579$ & $(134, 133)$ \\
\end{tabular}
\end{ruledtabular}
\end{table}

\textit{Result 2: LC5.}---Now we consider the five-qubit case
\begin{equation*}
\ket{LC_5}=CZ_{AB}CZ_{BC}CZ_{CD}CZ_{DE}\ket{+}^{\otimes 5},
\end{equation*}
with \(\mathcal E=ADE\), the LC4 measurements on $A,B,C,D$, and $E_0=Z_E,E_1=X_E$ (Fig.~\ref{fig:witnesses}(b)).
As in LC4, the construction combines a \(B\)-side distributed CHSH block with two \(C\)-side stabilizer locks.
The \(Z_B\)-branch stabilizers \(X_AZ_B\) and \(X_AZ_BZ_DX_E\) select \(A_0(1+D_1E_1)/2\), which equals \(A_0\) when \(D_1=E_1\) but vanishes when \(D_1=-E_1\).  We complete this sign on every record by defining
\begin{equation*}
\widetilde A_0=\frac{A_0(1+D_1E_1)-D_1+E_1}{2},
\qquad
\widetilde A_1=A_1D_0E_0.
\end{equation*}
Thus \(\widetilde A_0=A_0\) when \(D_1=E_1\) and \(\widetilde A_0=E_1\) otherwise; \(Z_AX_BX_DZ_E\) selects \(\widetilde A_1\).  These dichotomic early-side outcomes form the \(B\)-side block
\begin{equation*}
R_5:=2\avg{\widetilde A_0(B_0+B_1)
                 +\widetilde A_1(B_0-B_1)}.
\end{equation*}
The remaining stabilizers form two \(C\)-side locks,
\begin{align*}
L_{5,1}&:=2\left(\avg{A_0C_1D_1}
                 +\avg{A_0C_1E_1}-\avg{D_1E_1}\right),\\
L_{5,2}&:=2\avg{C_0D_0E_0}.
\end{align*}
The cluster state gives \((R_5,L_{5,1},L_{5,2})=(4\sqrt2,2,2)\).  The HIC projection contains three certified boundary points \((8,2,0)\), \((4,2,2)\), and \((8,-2,2)\).  Their edge directions share the primitive normal \((1,1,2)\), fixing the block weights.  An integer Farkas certificate proves
\(R_5+L_{5,1}+2L_{5,2}\le10\), and all three points attain the bound.  Explicit realizations are given in the Supplemental Material, Sec.~II~C.

Expanding this three-block inequality, for every $P\in\NS\cap\CL(BC|ADE)$, we have
\begin{equation}
\begin{aligned}
S_5={}&
\avg{A_0B_0}+\avg{A_0B_1}
-\avg{B_0D_1}-\avg{B_1D_1}\\
&+\avg{B_0E_1}+\avg{B_1E_1}
+2\avg{A_0C_1D_1}\\
&+2\avg{A_0C_1E_1}
-2\avg{D_1E_1}\\
&+\avg{A_0B_0D_1E_1}
+\avg{A_0B_1D_1E_1}\\
&+2\avg{A_1B_0D_0E_0}
-2\avg{A_1B_1D_0E_0}\\
&+4\avg{C_0D_0E_0}
\le 10 ,
\end{aligned}
\label{eq:S5-main}
\end{equation} which, as before, coincides with the fully local bound (achieved e.g.~by $A_0=B_0=B_1=C_0=+1$, any $A_1$, $C_1=D_1=E_1$, $D_0=E_0$). Supplemental Material Fig.~S1 provides a visual illustration. The four added terms \(-\avg{B_0D_1}-\avg{B_1D_1}+\avg{B_0E_1}+\avg{B_1E_1}\) vanish on \(\ket{LC_5}\), but change the supporting hyperplane: with them the saturating face has dimension \(133\) and is facet-defining, whereas without them its dimension is \(125\) (Supplemental Material, Sec.~VII~D). All supports lie in $ABDE$, $ACDE$, or lower marginals; none contains both $B$ and $C$, and
\begin{align*}
S_5^Q&=6+4\sqrt2\approx11.6569,\\
\nu_5^*&=\frac{10}{S_5^Q}=\frac{10}{6+4\sqrt2}\approx0.8579.
\end{align*} 
The spacetime configuration for this scenario is given in Supplemental Material, Sec.~I~E. Notice that this scenario could also be adapted in the same spacetime configuration as that of Ref.~\cite{Bancal2012} with $D$ and $E$ on the same side. Indeed, D and E could be seen as a single system undergoing two measurements $M_0=(D_0,E_0)$ and $M_1=(D_1,E_1)$ with four outcomes each, which also has the same HIC bound 10.

\textit{Tightness diagnostics.}---The block projections above show that the bounds $6$ and $10$ are attained: each supporting plane touches an HIC point. Thus the bounds are tight, but this does not yet show that either inequality is a facet of its full no-\(BC\) projected HIC polytope. Facet tightness requires the saturating points in no-\(BC\) correlator space to span a codimension-one face, as in the polyhedral treatment of Bell inequalities~\cite{Pironio2005,Ziegler1995}. This check is nontrivial because deterministic $B,C$ responses do not determine the projected vertices: their weights must additionally satisfy the no-signaling constraints. We therefore use the integer dual certificates to prove the bounds and identify which hidden-influence components can contribute at saturation. From these components, we construct feasible HIC points, project them into no-\(BC\) correlator space, and compute their affine spans by exact rational arithmetic. The active faces have dimensions $43$ for LC4 and $133$ for LC5, one below the corresponding polytope dimensions $44$ and $134$, so both inequalities are facet-defining. The certificates and their exact verification are summarized in Supplemental Material, Secs.~IV-V, and Table~\ref{tab:main-numbers}.

\textit{Discussion and outlook.}---We have shown that linear-cluster stabilizers yield compact, facet-defining witnesses for the projected no-\(BC\) HIC polytopes, with exact support bounds. Their ideal white-noise critical visibilities are \(0.8787\) for LC4 and \(0.8579\) for LC5 (Table~\ref{tab:main-numbers}), improving on the optimized value \(0.9526\) of the earlier finite-speed construction~\cite{Bancal2012}. 
This visibility comparison between LC4 and LC5 does not by itself determine experimental performance. 
Higher-body correlators are generally more sensitive to local readout errors. LC4 uses six two- and three-body correlators and no four-body term, whereas LC5 uses fourteen terms, including four-body \(ABDE\) correlators. LC4 is therefore the shorter witness and avoids four-body measurements, whereas LC5 offers the lower ideal white-noise threshold. The quoted visibilities are thus theory-level benchmarks rather than complete experimental thresholds. 
A realistic comparison must also account for finite-sample uncertainty, readout calibration, crosstalk, and shared-marginal inconsistencies.

The projected-polytope viewpoint raises two concrete questions. First, how does the optimal visibility scale with the number of parties, measurement settings, and outcomes? The dependence can be structural: in the conditioned tripartite binary-input, binary-output scenario, the projected HIC and no-signaling sets coincide, whereas the reported quantum separation was obtained only after giving the middle party three inputs and three outputs~\cite{Barnea2013}. Together with the four-party construction of Ref.~\cite{Bancal2012} and the present LC4 and LC5 witnesses, these results provide examples from three to five parties, but not yet a scaling behavior because the target states and measurement settings differ. 
Optimizing the visibility in each fixed setting under a common noise model would reveal whether additional parties or richer local measurements offer higher experimental robustness. 
Second, can the full facet structure of the projected no-\(BC\) HIC polytopes be characterized? Our certificates establish two facets, not complete facet lists. Complete facet descriptions would identify the most noise-robust no-\(BC\) linear witness. Exact enumeration in tractable cases and symmetry-reduced separation in larger ones could therefore reveal larger quantum-to-HIC gaps, or certify the present witnesses as optimal for their specified cluster-state correlations.

\vspace{.2cm}\noindent
\textit{Acknowledgements.}---The authors thank Jordi Tura, Anatoly Kulikov, Owidiusz Makuta, Jin-Fu Chen, and Yingjian Liu for helpful discussions. W.L.~and D.L.~are supported by the National Natural Science Foundation of China (T2225008), the Quantum Science and Technology-National Science and Technology Major Project (2021ZD0302203), the Fundamental and Interdisciplinary Disciplines Breakthrough Plan of the Ministry of Education of China (JYB2025XDXM112), the Shanghai Qi Zhi Institute Innovation Program (SQZ202318), and the Tsinghua University Dushi Program. M.H. and V.S. are supported by the National Research Foundation, Singapore through the National Quantum Office, hosted in A*STAR, under its Centre for Quantum Technologies Funding Initiative (S24Q2d0009).
The code for generating and verifying the certificates in this work was developed with the assistance of GPT-5.6 Sol.

\vspace{.2cm}\noindent
\textit{Data and Code availability.}---The certificate files and verifiers will be released upon publication. 

\bibliography{references}

\end{document}


\title{\texorpdfstring{Supplemental Material:\\ Cluster-State Witnesses of Finite-Speed Hidden Influences}{Supplemental Material: Cluster-State Witnesses of Finite-Speed Hidden Influences}}

\maketitle
\tableofcontents

\section{Finite-speed model, no-signaling, and the no-\texorpdfstring{\(BC\)}{BC} projection}
\label{sec:supp-model}

Throughout this supplement, LC4 and LC5 denote the four- and five-qubit linear-cluster witnesses, respectively.  We define the HIC set and its no-\(BC\) projection, derive both witnesses, and prove their support bounds and facet-defining properties.  The certification scope is the same as in the main text: the inequalities are certified for the no-\(BC\) projection \(\mathcal P_{\mathcal E}^{\mathrm{no}\text{-}BC}=\Pi_{\mathrm{no}\text{-}BC}(\NS\cap\CL(BC|\mathcal E))\), not for the ordinary Bell-local polytope and not for the full no-signaling set.

\subsection{Spacetime assumption}
The finite-speed hidden-influence model assumes a preferred inertial frame and a finite propagation speed $v>c$ for hidden causal influences.  If an event $\mathcal K_2$ lies inside the future $v$-cone of an event $\mathcal K_1$, hidden information generated at $\mathcal K_1$ may affect the outcome at $\mathcal K_2$.  If two events are outside each other's $v$-cones, no direct hidden influence can pass between them during that run.  The constraint is weaker than Bell locality because it permits hidden communication at \(v>c\), while requiring the observed probabilities to remain no-signaling.  The speed \(v\) characterizes only the hidden mechanism and does not modify the target quantum predictions.

Following Bancal \textit{et al.}, the relevant spacetime setting is multipartite~\cite{Bancal2012}.  Measurements on an early side \(\mathcal E\) occur before those at the two late parties \(B\) and \(C\), and the early events lie in the relevant past \(v\)-cones of the late events.  The two late events are arranged so that $B\sim C$, meaning that each is outside the other's $v$-cone.  The finite-speed model may therefore correlate the late parties with the early events, but it cannot use a hidden signal from $B$ to $C$ or from $C$ to $B$ once \((s_{\mathcal E},o_{\mathcal E})\) is fixed.

For four parties we use $\mathcal E=AD$. For five parties we use $\mathcal E=ADE$. The early-side settings and outcomes are denoted by $s_{\mathcal E}$ and $o_{\mathcal E}$. In the $B\sim C$ geometry the model imposes
\begin{equation}
P(o_{\mathcal E},b,c|s_{\mathcal E},y,z)
=\sum_{\lambda}\tau^{s_{\mathcal E}}_{o_{\mathcal E},\lambda}
P(b|y,\lambda)P(c|z,\lambda).
\label{eq:supp-cond-local}
\end{equation}
Here $\tau^{s_{\mathcal E}}_{o_{\mathcal E},\lambda}\ge0$. Summing Eq.~(\ref{eq:supp-cond-local}) over $b,c$ gives the early-event weight, and no-signaling identifies it with $P(o_{\mathcal E}|s_{\mathcal E})$; hence all weights vanish for null early events.
Only $B$ and $C$ factorize; the set is therefore not the fully local Bell polytope.

\begin{definition}[Full-distribution HIC set]
For a fixed early side \(\mathcal E\), the full-distribution HIC set is
\begin{equation}
\HIC_{\mathcal E}=\NS\cap\CL(BC|\mathcal E),
\label{eq:supp-hic-set}
\end{equation}
where $\NS$ denotes the set of full distributions satisfying operational no-signaling and $\CL(BC|\mathcal E)$ is the set of distributions admitting the subnormalized conditional-local decomposition in Eq.~(\ref{eq:supp-cond-local}).
\end{definition}

Every distribution generated by a finite-speed model satisfying the stated spacetime premise, conditional \(BC\) locality, and operational no-signaling lies in this set.  These conditions are necessary for the stated model class but do not characterize all finite-speed theories.

\begin{definition}[Projected no-\(BC\) hidden-influence set]
Let $\Pi_{\mathrm{no}\text{-}BC}$ discard every correlator coordinate containing both blind parties $B$ and $C$.  The polytope used for witness search, support certification, and facet verification is
\begin{equation}
\mathcal P_{\mathcal E}^{\mathrm{no}\text{-}BC}
:=\Pi_{\mathrm{no}\text{-}BC}(\HIC_{\mathcal E})
=\Pi_{\mathrm{no}\text{-}BC}\bigl(\NS\cap\CL(BC|\mathcal E)\bigr).
\label{eq:supp-projected-hic-set}
\end{equation}
Thus no-signaling and conditional locality are imposed in the full probability table before the no-\(BC\) coordinates are retained. This is not equivalent to intersecting the two separately projected sets $\Pi_{\mathrm{no}\text{-}BC}(\NS)\cap\Pi_{\mathrm{no}\text{-}BC}(\CL(BC|\mathcal E))$.
\end{definition}

For any party subset \(R\) with complement \(\bar R\), operational no-signaling requires that
\begin{equation}
\sum_{o_{\bar R}}P(o_R,o_{\bar R}|s_R,s_{\bar R})
\end{equation}
is independent of $s_{\bar R}$.  These identities are imposed together with normalization and nonnegativity.

\subsection{No-\texorpdfstring{\(BC\)}{BC} support and measured marginals}
The no-\(BC\) condition applies to the measured functional rather than to the prepared graph state.  The graph may therefore contain a \(BC\) edge although no witness term contains both parties.

\begin{definition}[no-\(BC\) correlator]
In a correlator expression, let $\operatorname{supp}(T)$ be the set of parties appearing in a term $T$.  The term is no-\(BC\) if $\{B,C\}\nsubseteq \operatorname{supp}(T)$.  A linear witness is no-\(BC\) if every nonzero term is no-\(BC\).
\end{definition}

With two settings per party, each party contributes three local factors: the identity and the observables for settings \(0\) and \(1\).  The identity coordinate is excluded.  For four parties there are $3^4-1=80$ nontrivial correlators.  The forbidden coordinates contain both $B$ and $C$ and arbitrary choices on the other two parties, giving $4\cdot 3^2=36$ forbidden coordinates.  Thus the four-party no-\(BC\) projected space has
\begin{equation}
3^4-1-4\cdot 3^2=44
\label{eq:supp-nobc-four}
\end{equation}
coordinates.  For five parties the corresponding number is
\begin{equation}
3^5-1-4\cdot 3^3=134 .
\label{eq:supp-nobc-five}
\end{equation}

Inspection of Eqs.~(\ref{eq:supp-S4}) and (\ref{eq:supp-S5}) shows that LC4 uses only the supports \(AB\), \(ABD\), \(CD\), and \(ACD\), whereas LC5 uses \(AB\), \(BD\), \(BE\), \(ACD\), \(ACE\), \(DE\), \(ABDE\), and \(CDE\).  None of these supports contains both \(B\) and \(C\).

\subsection{Relation to Bancal \textit{et al.}'s construction}
The original finite-speed argument can be organized as a conditional Clauser-Horne proof~\cite{Bancal2012,ClauserHorne1974}.  One fixes an early-side event, applies a bipartite local inequality to the conditional distribution of $B,C$, removes denominators by multiplying by the early-event probability, and then uses no-signaling to rewrite the final inequality in an observable form.  This route makes the causal assumptions explicit but does not by itself optimize the quantum-to-HIC ratio.  In the original work, the displayed witness has a quantum value \(7.2014\) for a bound \(7\), corresponding to only a \(2.9\%\) gap relative to the bound and an ideal white-noise threshold \(0.9720\).  The optimized value \(7.3481\) corresponds to a threshold of \(0.9526\).  We instead seek sparse no-\(BC\) witnesses with larger cluster-state violations relative to their HIC bounds.

Following the projected-polytope approach of Ref.~\cite{Barnea2013}, we choose stabilizer-based functionals directly in the projected coordinates and optimize them over full no-signaling extensions satisfying conditional \(BC\) locality.  Integer Farkas certificates give the support bounds, explicit primal points prove attainment, and exact rank and relative-interior feasibility calculations establish the facet-defining property.  Although \(BC\) does not appear in the final functionals, the discarded coordinates remain constrained by the full extensions.

\subsection{Operational marginal interpretation}
The witness is evaluated from marginals involving either \(B\) or \(C\) together with the early parties, all taken from a single operational distribution.  All correlators must be estimated from data compatible with a common \(B\sim C\) geometry and a fixed setting convention. If separate acquisition blocks are used, their shared marginals and source statistics must remain stable. We assume measurement independence and no postselection.

LC4 uses the \(ABD\) and \(ACD\) marginals, whereas LC5 uses \(ABDE\) and \(ACDE\); all lower-order terms are marginals of these families.

\subsection{A five-site spacetime arrangement for LC5}
\label{sec:supp-lc5-spacetime}
We now give an LC5 spacetime arrangement with five separate parties.  For each causally possible early sender, the other four records can be collected before light from that sender arrives.  This is the geometric condition needed to turn signaling into faster-than-light communication.  A finite experiment would also require repeated trials and a statistical decision rule. 

\begin{proposition}[Five-site LC5 arrangement]
Fix a finite hidden-influence speed \(v>c\), write \({\kappa}=v/c\), and use the time coordinate \(\theta=ct_{\rm lab}\).  We express \(\theta\) and all spatial coordinates in a common arbitrary unit of length.  Define
\begin{equation}
q:=\frac{1+{\kappa}^{-1}}{2},\qquad
{\zeta_*}:=\min\left\{
\frac{3}{22}\sqrt{({\kappa}q)^2-1},
\frac{3(1-q)}{4q}
\right\}.
\label{eq:supp-lc5-spacetime-parameters}
\end{equation}
For any \(0<{\zeta}<{\zeta_*}\), write \(\bm p=(p_{\parallel},p_{\perp})\) and place the five events at \(\mathcal K_i=(\theta_i;\bm p_i)\), with
\begin{equation}
\begin{aligned}
\mathcal K_A&=(0;0,0),&
\mathcal K_D&=(q;1,{\zeta}),&
\mathcal K_E&=(2q;2,4{\zeta}),\\
\mathcal K_B&=(5q;5,26{\zeta}),&
\mathcal K_C&=(5q;5,24{\zeta}).
\end{aligned}
\label{eq:supp-lc5-spacetime-coordinates}
\end{equation}
For event labels \(i,j\), write \(i\prec_v j\) when \(\theta_j>\theta_i\) and \(\lvert\bm p_j-\bm p_i\rvert<\kappa(\theta_j-\theta_i)\).  All five events are pairwise \(c\)-spacelike.  In the preferred frame they obey
\begin{equation}
A\prec_v D\prec_v E\prec_v B,C,
\qquad B\sim C.
\label{eq:supp-lc5-spacetime-order}
\end{equation}
For an event \(\mathcal K_i=(\theta_i;\bm p_i)\), define its ordinary future light cone by
\begin{equation*}
J_c^+(\mathcal K_i):=
\left\{(\theta;\bm p):\theta-\theta_i\geq
\left|\bm p-\bm p_i\right|\right\}.
\end{equation*}
Moreover, for every possible early source \(X\in\{A,D,E\}\), there is a collection event
\begin{equation}
\mathcal K_X^{\mathrm{col}}\in
\left(
\bigcap_{\substack{Y\in\{A,B,C,D,E\}\\Y\ne X}}
J_c^+(\mathcal K_Y)
\right)
\setminus J_c^+(\mathcal K_X).
\label{eq:supp-lc5-collection-region}
\end{equation}
At \(\mathcal K_X^{\mathrm{col}}\), the other four records have arrived, but light from \(X\) has not.
\end{proposition}

\begin{proof}
Because \({\kappa}>1\), we have \({\kappa}^{-1}<q<1\) and \({\kappa}q=({\kappa}+1)/2>1\).  For two sites with different \(p_{\parallel}\)-coordinates, the spatial separation is at least \(\lvert\Delta p_{\parallel}\rvert\), while the time separation is \(q\lvert\Delta p_{\parallel}\rvert\).  The only remaining pair, \(B,C\), is simultaneous and has separation \(2{\zeta}\).  Thus every pair is \(c\)-spacelike.

Let \(\mu:=\sqrt{({\kappa}q)^2-1}>0\).  For a link with \(\Delta\theta=q\Delta p_{\parallel}\), the \(v\)-cone condition is \(\lvert\Delta p_{\perp}\rvert<\mu\Delta p_{\parallel}\).  The four required links \(AD,DE,EB,EC\) have \((\Delta p_{\parallel},\lvert\Delta p_{\perp}\rvert)\) equal to \((1,{\zeta})\), \((1,3{\zeta})\), \((3,22{\zeta})\), and \((3,20{\zeta})\).  The first bound on \({\zeta}\) therefore gives
\begin{equation}
{\zeta}<\mu,\qquad 3{\zeta}<\mu,
\qquad 22{\zeta}<3\mu,
\qquad 20{\zeta}<3\mu.
\end{equation}
Hence \(A\prec_vD\prec_vE\prec_vB,C\).  The simultaneous and distinct events \(B,C\) are outside each other's \(v\)-cones, so \(B\sim C\).

We next prove Eq.~(\ref{eq:supp-lc5-collection-region}).  For \(s=0,1,2\), write \(\mathcal K_{X_s}=(qs;s,{\zeta}s^2)\), where \(X_0=A\), \(X_1=D\), and \(X_2=E\).  For each \(k\in\{0,1,2\}\), choose a unit vector \(\bm n_k=(n_{k,\parallel},n_{k,\perp})\) satisfying
\begin{equation}
n_{k,\parallel}+2k{\zeta} n_{k,\perp}=q,
\qquad n_{k,\perp}>0.
\label{eq:supp-lc5-retarded-direction}
\end{equation}
The line in Eq.~(\ref{eq:supp-lc5-retarded-direction}) intersects the unit circle because \(q<1\leq\sqrt{1+4k^2{\zeta}^2}\).  We choose the upper intersection, which has \(n_{k,\perp}>0\).  Define the arrival offset \(h_i(\bm n):=\theta_i-\bm n\cdot\bm p_i\).  It is the leading event-dependent part of the light-arrival time at a distant point in direction \(\bm n\).  Since \(n_{k,\perp}>0\), we also have \(h_C(\bm n_k)>h_B(\bm n_k)\). Equation~(\ref{eq:supp-lc5-retarded-direction}) gives, for \(s\ne k\),
\begin{align}
{h_{X_k}}(\bm n_k)-{h_{X_s}}(\bm n_k)
&={\zeta} n_{k,\perp}(s-k)^2>0,\\
{h_{X_k}}(\bm n_k)-\max\{h_B(\bm n_k),h_C(\bm n_k)\}
&={\zeta} n_{k,\perp}\bigl[(5-k)^2-1\bigr]>0.
\label{eq:supp-lc5-retarded-gaps}
\end{align}
Thus the source \(X_k\) has a larger arrival offset than each of the other four events.
At the spatial point \(R\bm n_k\), light from \(\mathcal K_i\) arrives at
\begin{equation}
\theta_i+|R\bm n_k-\bm p_i|
=R+h_i(\bm n_k)+\mathcal O(R^{-1}).
\end{equation}
There are only five events, and all gaps in Eq.~(\ref{eq:supp-lc5-retarded-gaps}) are strict.  We can therefore choose a finite \(R\) for which light from \(X_k\) arrives last.  Let \(i\) range over \(\{A,B,C,D,E\}\), and choose \(\theta_k^{\mathrm{col}}\) such that
\begin{equation*}
\max_{i\ne X_k}
\left\{\theta_i+\left|R\bm n_k-\bm p_i\right|\right\}
<\theta_k^{\mathrm{col}}
<\theta_{X_k}+\left|R\bm n_k-\bm p_{X_k}\right|.
\end{equation*}
Then \(\mathcal K_{X_k}^{\mathrm{col}}=(\theta_k^{\mathrm{col}};R\bm n_k)\) lies inside the other four future light cones and outside that of \(X_k\).  This proves Eq.~(\ref{eq:supp-lc5-collection-region}).
\end{proof}

We now obtain the two no-\(BC\) marginal families.  Set
\begin{equation}
\theta_{\rm d}:=2q{\zeta}.
\end{equation}
Since \(|\bm p_B-\bm p_C|=2{\zeta}\) and \({\kappa}^{-1}<q<1\),
\begin{equation}
\frac{|\bm p_B-\bm p_C|}{{\kappa}}<\theta_{\rm d}
<|\bm p_B-\bm p_C|.
\label{eq:supp-lc5-timing-window}
\end{equation}
A delay of \(C\) replaces \(\mathcal K_C\) by \(\mathcal K_C'=(5q+\theta_{\rm d};5,24{\zeta})\) and leaves the other events fixed.  A delay of \(B\) is defined in the same way.  The two inequalities in Eq.~(\ref{eq:supp-lc5-timing-window}) follow directly from
\begin{equation*}
{\kappa}\theta_{\rm d}-2{\zeta}={\zeta}({\kappa}-1)>0,
\qquad
2{\zeta}-\theta_{\rm d}=\frac{{\zeta}({\kappa}-1)}{{\kappa}}>0.
\end{equation*}
The first quantity makes the delayed late pair \(v\)-connected, while the second keeps it \(c\)-spacelike.  Thus delaying \(C\) gives \(A\prec_vD\prec_vE\prec_vB\prec_vC'\).  Delaying \(B\) instead gives \(A\prec_vD\prec_vE\prec_vC\prec_vB'\).

The second bound on \({\zeta}\) gives \(\theta_{\rm d}<3(1-q)/2<3(1-q)\).  For an early event \(X_s\), the longitudinal gap to either late site is \(5-s\geq3\).  Hence \(q(5-s)+\theta_{\rm d}<5-s\), so the delayed event remains \(c\)-spacelike from \(A,D,E\).

\begin{lemma}[The delay preserves the required marginal]
Let \(G_0\) denote the target geometry with \(B\sim C\).  Let \(G_C\) denote the branch in which only \(C\) is delayed, and define \(G_B\) analogously.  As in Ref.~\cite{Bancal2012}, suppose that each delay choice is made locally at the original late event.  The choice is random and independent of the hidden variables, settings, and earlier outcomes.  Then, for every \((x,y,z,w,u)\),
\begin{align}
P_{G_0}(a,b,d,e\mid x,y,z,w,u)
&=P_{G_C}(a,b,d,e\mid x,y,z,w,u),\\
P_{G_0}(a,c,d,e\mid x,y,z,w,u)
&=P_{G_B}(a,c,d,e\mid x,y,z,w,u).
\end{align}
\end{lemma}
\begin{proof}
Consider the choice made at \(C\).  The events \(A,D,E\) have already occurred, while \(B\) is simultaneous with \(C\) and outside its \(v\)-cone.  The choice at \(C\) can therefore affect none of the records \(A,B,D,E\).  Its independence also prevents the choice from selecting a different hidden-variable subensemble.  This proves the first equality.  Exchanging \(B\) and \(C\) proves the second.
\end{proof}

Assume that a finite-\(v\) model gives the quantum predictions whenever all five events form a \(v\)-connected chain.  Both \(G_C\) and \(G_B\) are such chains.  The lemma therefore transfers their quantum \(ABDE\) and \(ACDE\) marginals to the target geometry \(G_0\).  Every term in \(S_5\) belongs to one of these marginals.  Hence
\begin{equation}
S_5=6+4\sqrt2>10.
\end{equation}
Because \(B\sim C\), the \(BC|ADE\) correlations are local.  If the target distribution were also no-signaling, the certified bound \(S_5\leq10\) would apply. Thus, under the stated reproduction assumption, the target distribution in \(G_0\) must violate operational no-signaling.

In the forward \(v\)-causal order, this signaling can originate only from \(A\), \(D\), or \(E\).  The late parties can affect neither earlier events nor each other.  If a dependence involves several early settings, vary them one at a time.  At least one single-party change must alter the later distribution.  Equation~(\ref{eq:supp-lc5-collection-region}) supplies a collection event outside that party's future light cone.  This proposition establishes the required geometry.  A finite-data test must additionally specify repeated trials and an error-controlled decision rule.

The transverse coordinate is essential.  To see why, consider a strictly \(1+1\)-dimensional layout whose measurement and collection events lie on the same spatial line.  Label the three early sites by their positions, \(\chi_L<\chi_M<\chi_R\), and let \(\mathcal T_i(\chi)=\theta_i+|\chi-\chi_i|\) be the light-arrival time at \(\chi\).  Pairwise \(c\)-separation gives
\begin{equation}
\begin{aligned}
\mathcal T_L(\chi)-\mathcal T_M(\chi)&=\theta_L-\theta_M+\chi_M-\chi_L>0,
&&\chi\geq \chi_M,\\
\mathcal T_R(\chi)-\mathcal T_M(\chi)&=\theta_R-\theta_M+\chi_R-\chi_M>0,
&&\chi\leq \chi_M.
\end{aligned}
\end{equation}
Thus light from the middle site is never the last to arrive anywhere on that line.  A strictly \(1+1\)-dimensional layout therefore misses one possible sender.  The two-dimensional layout in Fig.~\ref{fig:supp-lc5}(b) covers all three. The construction works for every finite \({\kappa}>1\), but the allowed range of \({\zeta}\) becomes small as \({\kappa}\to1^+\).

\subsection{Coordinate conventions}
The calculations use bit outputs internally, while the manuscript uses \(\pm1\) observables.  The conversion is the usual parity transform.  For a party subset $R$, a bit-output probability distribution gives the correlator
\begin{equation}
\Gamma_R(\bm{s})=\sum_{\bm{o}}(-1)^{\sum_{i\in R}o_i}P(\bm{o}|\bm{s}).
\end{equation}
Linear inequalities can be expressed either in probability coordinates or in correlator coordinates.  The present witness coefficients are stated in correlator form because this makes the stabilizer derivation transparent.  The LP certificate is checked after pulling those correlator coefficients back to the deterministic hidden-influence variables.

This convention also fixes the sign of the rotated $B$ observables.  We take
\begin{equation}
B_0=\frac{Z+X}{\sqrt2},\qquad B_1=\frac{Z-X}{\sqrt2},
\end{equation}
so that $B_0+B_1=\sqrt2 Z$ and $B_0-B_1=\sqrt2 X$.  With this choice, the LC4 term $-\avg{A_1B_1D_0}$ contributes positively because the corresponding cluster-state expectation value is negative.  We neither postselect records nor relabel physical outcomes.  The variable \(\widetilde A_0\) introduced below is only a deterministic function of recorded outcomes, and its full correlator expansion is retained in the witness.

\section{Derivation of the witness coefficients}
\label{sec:supp-construction}

\subsection{Block-coordinate construction}
Each witness consists of a distributed CHSH block on the \(B\) side and one or more \(B\)-free stabilizer blocks.  A supporting face of the projected HIC set determines the relative block coefficients through a primitive integer normal, and an integer Farkas certificate proves the resulting bound over the full projected HIC set.  The stabilizers determine the cluster-state value.  This construction does not claim global optimality in the \(44\)- or \(134\)-dimensional projected spaces.

Here ``distributed CHSH'' refers only to the algebraic form.  Because early outcomes may influence \(B\), the block alone need not obey the usual Bell-local CHSH bound.  Graph-state Bell methods, including Ref.~\cite{CoiteuxRoy2025}, motivate the construction, but the relevant set is the HIC set in Eq.~(\ref{eq:supp-hic-set}), rather than the alternative locality models considered there.

\subsection{LC4: a two-dimensional support problem}
Define

\begin{align}
R_4&:=\avg{A_0(B_0+B_1)}
      +\avg{A_1D_0(B_0-B_1)},
\label{eq:supp-R4-definition}\\
L_4&:=\avg{C_0D_0}+\avg{A_0C_1D_1}.
\label{eq:supp-L4-definition}
\end{align}

The variables \(U_0=A_0\) and \(U_1=A_1D_0\) are both dichotomic, so \(R_4=\avg{U_0(B_0+B_1)+U_1(B_0-B_1)}\) is a CHSH polynomial between \(B\) and an effective early-side party.  The two terms in \(L_4\) form the stabilizer block.  At the LC4 cluster-state point,

\begin{equation}
R_4^Q=2\sqrt2,\qquad L_4^Q=2,
\label{eq:supp-lc4-block-qpoint}
\end{equation}
because
\begin{equation}
A_0(B_0+B_1)=\sqrt2 K_A,\quad
A_1D_0(B_0-B_1)=\sqrt2 K_BK_D,\quad
C_0D_0=K_D,\quad A_0C_1D_1=K_AK_C.
\end{equation}
Under projection onto \((R_4,L_4)\), the upper HIC boundary contains the exact feasible points
\begin{equation}
v_1=(4,1),\qquad v_2=(2,2).
\label{eq:supp-lc4-block-vertices}
\end{equation}
These points are realized by the following distributions.  With a shared fair bit \(\xi\), the response rule

\begin{equation}
a=\xi,\quad d=0,\quad b=\xi\oplus xy,\quad c=z
\label{eq:supp-lc4-v1-model}
\end{equation}
is no-signaling, conditionally \(BC\)-local, and gives \(v_1\).  The deterministic rule

\begin{equation}
a=x,\quad d=1,\quad b=0,\quad c=1
\label{eq:supp-lc4-v2-model}
\end{equation}
gives \(v_2\).  Here \(x,y,z,w\) are the settings of \(A,B,C,D\), respectively, and all displayed letters on the right-hand side are bits.
The difference \(v_2-v_1=(-2,1)\) has the primitive outward normal \((1,2)\), so the candidate supporting line is

\begin{equation}
R_4+2L_4=6.
\label{eq:supp-lc4-block-plane}
\end{equation}

The integer dual certificate in Sec.~\ref{sec:supp-certificates} proves that the entire projected HIC set lies below this line.  Together with the algebraic bounds \(R_4\le4\) and \(L_4\le2\), it leaves \(v_1,v_2\) as the only upper-boundary vertices relevant to nonnegative block weights.  For \(p,\ell\ge0\), these half-spaces determine the support explicitly.  If \(\ell\le2p\), then
\begin{equation}
pR_4+\ell L_4
=\frac{\ell}{2}(R_4+2L_4)
+\left(p-\frac{\ell}{2}\right)R_4
\le4p+\ell ,
\end{equation}
whereas for \(\ell\ge2p\),
\begin{equation}
pR_4+\ell L_4
=p(R_4+2L_4)+(\ell-2p)L_4
\le2p+2\ell .
\end{equation}
The two bounds are attained by \(v_1\) and \(v_2\), respectively. Hence the support function for nonnegative weights is

\begin{equation}
\varphi_4(p,\ell)
:=\max_{P\in\HIC_{AD}}\bigl(pR_4(P)+\ell L_4(P)\bigr)
=\max\{4p+\ell,\,2p+2\ell\},
\qquad p,\ell\ge0.
\label{eq:supp-lc4-block-support}
\end{equation}

Define \(\pi_4(P):=(R_4(P),L_4(P))\), and let \(\mathcal L\) denote the fully Bell-local set.  The complete two-dimensional projections shown in Fig.~2(d) of the main text are
{
\begin{align}
\pi_4(\mathcal L)
&=\{(R_4,L_4): |R_4|\le2,\ |L_4|\le2\},\nonumber\\
\pi_4(\HIC_{AD})
&=\{(R_4,L_4): |R_4|\le4,\ |L_4|\le2,\
|R_4|+2|L_4|\le6\},\nonumber\\
\pi_4(\NS)
&=\{(R_4,L_4): |R_4|\le4,\ |L_4|\le2\}.
\label{eq:supp-lc4-full-block-projections}
\end{align}
}
Thus the Bell-local projection is a square, the HIC projection an octagon, and the no-signaling projection a rectangle.  Independent global output flips at \(B\) and \(C\) generate the other three slanted HIC facets from \(R_4+2L_4\le6\), while \(Q=(2\sqrt2,2)\) lies on the no-signaling boundary and outside the HIC octagon.

The corresponding cluster-state value is \(2\sqrt2 p+2\ell\).
For \(p>0\), set \(\bar\ell:=\ell/p\ge0\).  Since
\((4+\bar\ell)-(2+2\bar\ell)=2-\bar\ell\), the first branch of
Eq.~(\ref{eq:supp-lc4-block-support}) is active for \(0\le \bar\ell\le2\),
whereas the second is active for \(\bar\ell\ge2\).  Hence the scale-independent ratio is
{
\begin{equation}
\rho_4(\bar\ell)
:=\frac{2\sqrt2+2\bar\ell}{\max\{4+\bar\ell,\,2+2\bar\ell\}}
=
\begin{cases}
\displaystyle\frac{2\sqrt2+2\bar\ell}{4+\bar\ell},&0\le \bar\ell\le2,\\[6pt]
\displaystyle\frac{\sqrt2+\bar\ell}{1+\bar\ell},&\bar\ell\ge2.
\end{cases}
\label{eq:supp-lc4-ratio}
\end{equation}
}
Away from the branch point, its derivative is
{
\begin{equation}
\rho_4'(\bar\ell)
=
\begin{cases}
\displaystyle\frac{8-2\sqrt2}{(4+\bar\ell)^2}>0,&0\le \bar\ell<2,\\[6pt]
\displaystyle\frac{1-\sqrt2}{(1+\bar\ell)^2}<0,&\bar\ell>2.
\end{cases}
\label{eq:supp-lc4-ratio-derivative}
\end{equation}
}
Hence \(\rho_4\) is strictly increasing up to \(\bar\ell=2\) and strictly
decreasing thereafter, so its unique global maximum is attained at
\(\bar\ell=2\), with
\(\rho_4(2)=(2+\sqrt2)/3\).  The remaining boundary ray
\(p=0,\ell>0\) gives ratio \(1<\rho_4(2)\), while
\((p,\ell)=(0,0)\) does not define a witness.  Thus the unique maximizing
ray is \(\ell=2p\).
Removing the irrelevant overall scale gives \((p,\ell)=(1,2)\), hence Eq.~(\ref{eq:supp-S4}).

\subsection{LC5: dichotomic completion and a three-block support problem}
The two stabilizers containing \(Z_B\) (\(X_AZ_B\) and \(X_AZ_BZ_DX_E\)) have the half-sum

\begin{equation}
\frac{X_AZ_B+X_AZ_BZ_DX_E}{2}
=X_AZ_B\frac{1+Z_DX_E}{2}.
\label{eq:supp-lc5-raw-sign}
\end{equation}

Under the chosen measurement labels, this half-sum suggests the candidate effective variable \(A_0(1+D_1E_1)/2\).  On the target state, \(D_1E_1\) measures \(K_E=Z_DX_E\), so this variable equals \(A_0\).  A generic HIC distribution may instead have \(D_1=-E_1\), for which the candidate vanishes and cannot serve as a dichotomic CHSH observable.  We therefore complete the first variable on that sector and take \(A_1D_0E_0\), the early-side factor of \(K_BK_D=Z_AX_BX_DZ_E\), as the second effective variable:

\begin{equation}
\widetilde A_0
=\frac{A_0(1+D_1E_1)-D_1+E_1}{2},
\qquad
\widetilde A_1=A_1D_0E_0.
\label{eq:supp-effective-A-lc5}
\end{equation}

Both variables are deterministic functions of recorded outcomes, not additional physical measurements.  If \(D_1=E_1\), then \(\widetilde A_0=A_0\); if \(D_1=-E_1\), then \(\widetilde A_0=E_1=-D_1\).  Hence both \(\widetilde A_0\) and \(\widetilde A_1\) are dichotomic on every record, and \(\widetilde A_0=A_0\) whenever \(D_1=E_1\).
Define the three blocks

\begin{align}
R_5&:=2\avg{\widetilde A_0(B_0+B_1)
                 +\widetilde A_1(B_0-B_1)},
\label{eq:supp-R5-definition}\\
L_{5,1}&:=2\left(\avg{A_0C_1D_1}
                  +\avg{A_0C_1E_1}
                  -\avg{D_1E_1}\right),
\label{eq:supp-L51-definition}\\
L_{5,2}&:=2\avg{C_0D_0E_0}.
\label{eq:supp-L52-definition}
\end{align}
Expanding the first definition gives

\begin{align}
R_5={}&\avg{A_0B_0}+\avg{A_0B_1}
-\avg{B_0D_1}-\avg{B_1D_1}
+\avg{B_0E_1}+\avg{B_1E_1}\nonumber\\
&+\avg{A_0B_0D_1E_1}+\avg{A_0B_1D_1E_1}
+2\avg{A_1B_0D_0E_0}-2\avg{A_1B_1D_0E_0}.
\label{eq:supp-R5-expanded}
\end{align}

The four two-body terms in Eq.~(\ref{eq:supp-R5-expanded}) arise from the dichotomic completion in Eq.~(\ref{eq:supp-effective-A-lc5}) and have zero expectation on \(\ket{LC_5}\).  We refer to them as the completion terms.

The block \(L_{5,1}\) obeys a pointwise bound on each joint \((A_0,C_1,D_1,E_1)\) record.  Set \(g_1:=A_0C_1D_1\) and \(g_2:=A_0C_1E_1\).  Then \(g_1g_2=D_1E_1\), and for \(g_1,g_2\in\{\pm1\}\),

\begin{equation}
g_1+g_2-g_1g_2\le1.
\label{eq:supp-triangle-lock}
\end{equation}
Hence \(L_{5,1}\le2\); similarly \(L_{5,2}\le2\), while the algebraic CHSH bound gives \(R_5\le8\).
The cluster-state block values are
\begin{equation}
R_5^Q=4\sqrt2,\qquad
L_{5,1}^Q=2,\qquad
L_{5,2}^Q=2.
\label{eq:supp-lc5-block-qpoint}
\end{equation}

On the target state, the terms of \(R_5\) with nonzero expectation combine into \(\sqrt2 K_A\), \(\sqrt2 K_AK_E\), and \(2\sqrt2 K_BK_D\).  The terms in \(L_{5,1}\) reduce to \(2K_AK_C\), \(2K_AK_CK_E\), and \(-2K_E\), while \(L_{5,2}=2K_D\).

In the three-block coordinates, the following HIC points lie on a common upper face:

\begin{equation}
w_1=(8,2,0),\qquad
w_2=(4,2,2),\qquad
w_3=(8,-2,2).
\label{eq:supp-lc5-block-vertices}
\end{equation}
These points are realized by the following distributions.  With all inputs and outputs written as bits and \(\xi\) a shared fair bit, take

\begin{align}
P_1:\quad&
a=1\oplus x,\quad d=\xi,\quad e=0,\quad c=z,\quad
b=\xi\oplus y\oplus w(1\oplus y),\nonumber\\
P_2:\quad&
a=1\oplus x,\quad d=0,\quad e=1,\quad c=1\oplus z,\quad b=1,\nonumber\\
P_3:\quad&
a=\xi,\quad d=1,\quad e=u,\quad c=1,\quad
b=\xi\oplus x(1\oplus y).
\label{eq:supp-lc5-vertex-models}
\end{align}

\begin{figure}[t]
\centering
\includegraphics[width=0.80\textwidth]{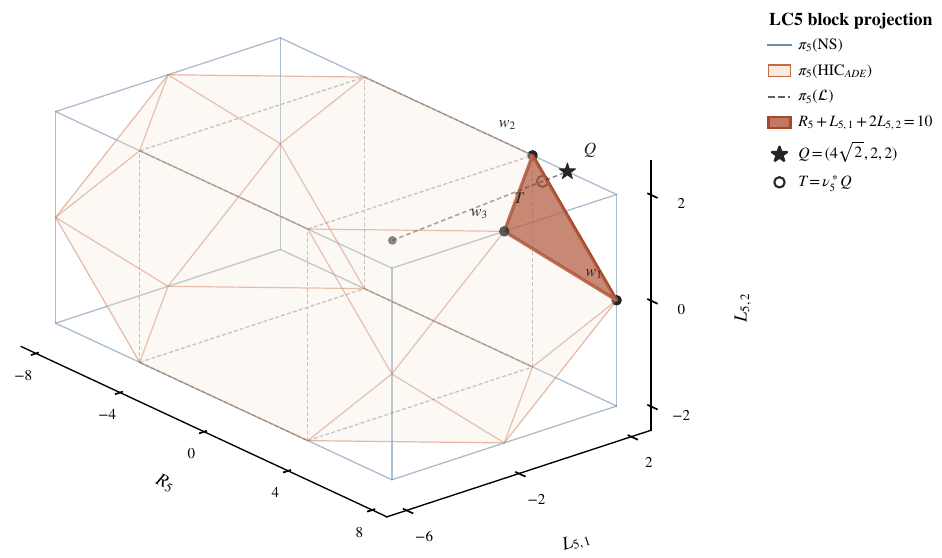}
\caption{\textbf{Exact LC5 three-block projection.}  For \(\pi_5(P):=(R_5(P),L_{5,1}(P),L_{5,2}(P))\), the blue wireframe, orange polytope, and dashed gray wireframe show \(\pi_5(\NS)\), \(\pi_5(\HIC_{ADE})\), and the Bell-local projection \(\pi_5(\mathcal L)\), respectively.  The highlighted triangular facet has vertices \(w_1,w_2,w_3\) from Eq.~(\ref{eq:supp-lc5-block-vertices}) and lies in the supporting plane \(R_5+L_{5,1}+2L_{5,2}=10\).  The cluster-state point \(Q_5=(4\sqrt2,2,2)\), labeled \(Q\) in the figure, lies in \(\pi_5(\NS)\) but outside \(\pi_5(\HIC_{ADE})\).  The white-noise ray from the origin meets the facet at \(T_5=\nu_5^*Q_5\), labeled \(T\) in the figure, where \(\nu_5^*=10/(6+4\sqrt2)\).}
\label{fig:supp-lc5-projection}
\end{figure}

Here \((x,y,z,w,u)\) are the settings of \((A,B,C,D,E)\), and direct substitution maps \(P_i\) to \(w_i\).  For \(P_1\), summing over \(d=\xi\) makes \(b\) uniform and removes its apparent dependence on \(w\).  For \(P_3\), summing over \(a=\xi\) removes the dependence on \(x\).  All remaining dependence on the settings is local.  Once the early settings and outcomes are fixed, \(\xi\) is fixed and \(B,C\) respond separately to \(y,z\).  Hence all three distributions are no-signaling and conditionally \(BC\)-local.

The two independent edge directions are \((-4,0,2)\) and \((0,-4,2)\).  Their primitive common normal is therefore \((1,1,2)\), giving the plane

\begin{equation}
R_5+L_{5,1}+2L_{5,2}=10.
\label{eq:supp-lc5-block-plane}
\end{equation}

The integer Farkas certificate proves that this plane supports the full projected HIC set.  Together with the three algebraic block bounds, it leaves \(w_1,w_2,w_3\) as the only upper-boundary vertices relevant to nonnegative block weights.  Since all three are attainable in the HIC set, the support for nonnegative block weights is exactly

\begin{equation}
\begin{split}
\varphi_5(p,\ell_1,\ell_2)
&:=\max_{P\in\HIC_{ADE}}
   \bigl(pR_5(P)+\ell_1L_{5,1}(P)+\ell_2L_{5,2}(P)\bigr)\\
&=\max\{8p+2\ell_1,\,4p+2\ell_1+2\ell_2,\,
8p-2\ell_1+2\ell_2\},
\qquad p,\ell_1,\ell_2\ge0.
\end{split}
\label{eq:supp-lc5-block-support}
\end{equation}

Writing \(\pi_5(P):=\bigl(R_5(P),L_{5,1}(P),L_{5,2}(P)\bigr)\), the exact HIC projection is
\[
\pi_5(\HIC_{ADE})
=
\pi_5(\NS)\cap
\left\{
(R,L_1,L_2):
|R|+|L_1+2|+2|L_2|\le12
\right\}.
\]
Fig.~\ref{fig:supp-lc5-projection} shows the nested block projections and highlights the supporting triangle through \(w_1,w_2,w_3\).

Each branch is attained by the corresponding point in Eq.~(\ref{eq:supp-lc5-block-vertices}).  Let \(b_1,b_2,b_3\) denote the three branches in Eq.~(\ref{eq:supp-lc5-block-support}).  The cluster-state numerator satisfies
{
\begin{equation}
4\sqrt2 p+2\ell_1+2\ell_2
=\frac{2(\sqrt2-1)}{5}b_1
+\frac{6-\sqrt2}{5}b_2
+\frac{\sqrt2-1}{5}b_3 .
\label{eq:supp-lc5-ratio-certificate}
\end{equation}
}
All three coefficients are positive.  Since \(b_i\le\varphi_5\), the ratio is at most \((3+2\sqrt2)/5\).  Equality for a nonzero weight vector requires \(b_1=b_2=b_3\), which gives \(\ell_1=p\) and \(\ell_2=2p\).  Thus the unique maximizing ray is \((p,\ell_1,\ell_2)\propto(1,1,2)\).  These weights yield Eq.~(\ref{eq:supp-S5}) and the coefficient pattern \(1,2,4\).

\section{Deterministic hidden-influence parametrization}
\label{sec:supp-param}

\subsection{Variables}
The blind pair has binary inputs $y,z\in\{0,1\}$ and binary outputs.  We use deterministic response functions
\begin{equation}
\beta,\gamma:\{0,1\}\to\{0,1\}.
\end{equation}
Refining \(\lambda\) into deterministic responses, write
\(t^{s_{\mathcal E}}_{o_{\mathcal E},\beta,\gamma}
:=\tau^{s_{\mathcal E}}_{o_{\mathcal E},(\beta,\gamma)}\).
There are four possible $\beta$ functions and four possible $\gamma$ functions.  In the four-party LC4 scenario the early settings are $(x,w)$ and the early outcomes are $(a,d)$, so the hidden-influence variables are
\begin{equation}
t^{xw}_{ad,\beta,\gamma}\ge0 .
\end{equation}
There are $2^2$ early settings, $2^2$ early outcomes, and $4^2$ deterministic response pairs, giving $256$ variables.  In the five-party LC5 scenario the variables are
\begin{equation}
t^{xwu}_{ade,\beta,\gamma}\ge0 ,
\end{equation}
giving $2^3\cdot2^3\cdot4^2=1024$ variables.
For a fixed early setting $s_{\mathcal E}$, the variables form a nonnegative measure over early outcomes and blind deterministic response functions.  Normalization requires
\begin{equation}
\sum_{o_{\mathcal E},\beta,\gamma}t^{s_{\mathcal E}}_{o_{\mathcal E},\beta,\gamma}=1
\label{eq:supp-normalization}
\end{equation}
for each $s_{\mathcal E}$.  The probabilities are reconstructed as
\begin{equation}
P(o_{\mathcal E},b,c|s_{\mathcal E},y,z)
=\sum_{\beta,\gamma}t^{s_{\mathcal E}}_{o_{\mathcal E},\beta,\gamma}
\mathbf{1}[b=\beta(y)]\mathbf{1}[c=\gamma(z)] .
\label{eq:supp-primal-map}
\end{equation}
Equation~(\ref{eq:supp-primal-map}) is the deterministic specialization of Eq.~(\ref{eq:supp-cond-local}); stochastic late responses are absorbed into convex mixtures of deterministic functions.

\subsection{No-signaling rows}
Normalization and no-signaling are collected into an equality system
\begin{equation}
\mathsf N t=r,\qquad t\ge0 .
\label{eq:supp-primal-feasible}
\end{equation}
The calculations use reduced equality systems with \(132\) rows for LC4 and \(776\) rows for LC5, instead of the corresponding full systems with \(260\) and \(1288\) rows. Exact rational row reduction shows that, in each scenario, the reduced and full coefficient matrices have the same row span, and so do their augmented matrices. Their ranks are \(64\) for LC4 and \(334\) for LC5. The reduced and full systems therefore define the same affine feasible set.

No-signaling under arbitrary changes of settings follows from the single-party setting-change equalities by successive setting changes and marginalization. In the deterministic-response parametrization, the rows associated with changes in the settings \(y\) and \(z\) of \(B\) and \(C\) are satisfied identically. Hence only setting changes at parties in the early side \(\mathcal E\) produce nontrivial constraints.

The no-signaling rows can be written directly in the probability representation and then pulled back through Eq.~(\ref{eq:supp-primal-map}).  For instance, in LC4 the \(BCD\) marginal must not depend on the setting \(x\) of \(A\):
{
\begin{equation}
\sum_aP(a,d,b,c|x,w,y,z)
=
\sum_aP(a,d,b,c|x',w,y,z)
\label{eq:supp-ns-row-example}
\end{equation}
}
for all choices of the displayed variables.  Equation~(\ref{eq:supp-ns-row-example}) therefore gives a linear constraint on the early-setting-dependent weights \(t\).

\subsection{Correlator map}
We encode each binary output \(o\in\{0,1\}\) as \({O}=(-1)^o\).  A correlator is
\begin{equation}
\avg{\prod_{i\in R}O_{i,s_i}}
=
\sum_{\bm{o}}(-1)^{\sum_{i\in R}o_i}
P(\bm{o}|\bm{s}) .
\end{equation}
Let \(\mathsf M\) map feasible weight vectors \(t\) to no-\(BC\) correlators.  For a witness coefficient vector \(\omega\), the objective in the \(t\)-coordinates is \(\alpha=\mathsf M^T\omega\).  The facets discussed in the main text are facets of this projected polytope, not of the higher-dimensional \(t\)-space.

\section{Farkas certificates and support bounds}
\label{sec:supp-certificates}

\subsection{Primal support program and dual certificates}
For a witness objective $S=\alpha^Tt$, the hidden-influence support value is obtained from the linear program
\begin{equation}
\mathcal B=\max\{\alpha^Tt:\mathsf N t=r,\ t\ge0\}.
\label{eq:supp-primal-lp}
\end{equation}
If \(\mathsf N\) has \(m\) rows, a vector \(\eta\in\mathbb R^m\), unrestricted in sign, is dual feasible when
\begin{equation}
\mathsf N^T\eta\ge \alpha.
\label{eq:supp-dual-feasible}
\end{equation}
If, in addition, $r^T\eta=\mathcal B$, then for every feasible point $t$,
\begin{equation}
\alpha^Tt\le \eta^T\mathsf N t=\eta^Tr=\mathcal B.
\end{equation}
Equivalently,
\begin{equation}
\mathcal B-S=(\mathsf N^T\eta-\alpha)^Tt\ge0 .
\label{eq:supp-farkas}
\end{equation}
A vector \(\eta\) satisfying Eq.~(\ref{eq:supp-dual-feasible}) and \(r^T\eta=\mathcal B\) is a Farkas certificate for the support bound~\cite{Schrijver1986}.  Equation~(\ref{eq:supp-farkas}) then proves the bound exactly.

\begin{proposition}[Exact HIC support values]
\label{prop:supp-support-bounds}
The support values of \(S_4\) on \(\mathcal P_{AD}^{\mathrm{no}\text{-}BC}\) and of \(S_5\) on \(\mathcal P_{ADE}^{\mathrm{no}\text{-}BC}\) are \(6\) and \(10\), respectively.
\end{proposition}

\begin{proof}
For LC4 and LC5, respectively, the selected integer dual vectors satisfy \(r^T\eta=6\) and \(10\), together with \(\mathsf N^T\eta-\alpha\ge0\) componentwise.  Among dual-optimal vectors satisfying \(\mathsf N^T\eta\ge\alpha\) and \(r^T\eta=\mathcal B\) in the reduced equality representation, the minimum norm \(\|\eta\|_1\) is \(70\) for LC4 and \(362\) for LC5.  The selected minimizers have \(22\) and \(98\) nonzero entries.  For each witness, define the dual slack \(\delta=\mathsf N^T\eta-\alpha\) and its zero-slack index set \(J=\{j:\delta_j=0\}\).  Both the norms and the nonzero counts depend on the chosen equality-row representation and are not invariants of the projected polytopes.  Substitution into Eq.~(\ref{eq:supp-farkas}) proves the two upper bounds for every feasible hidden-influence point.

The primal points certified in Sec.~\ref{sec:supp-facet} are supported on \(J\), so \(\delta^Tt=0\).  They therefore saturate Eq.~(\ref{eq:supp-farkas}), giving \(S_4=6\) and \(S_5=10\).
\end{proof}

\subsection{Exact certificate and facet verification}
For each witness, the integer dual vector satisfies \(r^T\eta=\mathcal B\) and \(\mathsf N^T\eta-\alpha\ge0\) componentwise.  Independently, the absence of coefficients involving both \(B\) and \(C\) verifies that the witness is no-\(BC\).  Floating-point optimization is used only to generate candidates; after integer or rational reconstruction, all stated identities, inequalities, and ranks are verified exactly.

Table~\ref{tab:supp-diagnostics} summarizes the support certificates and facet calculations for LC4 and LC5.

The ranks of \(\mathsf N\), \((\mathsf N;\mathsf M)\), \(\mathsf N_J\), and \((\mathsf N_J;\mathsf M_J)\) are evaluated over \(\mathbb Q\).  Here \(\mathsf N_J\) denotes restriction to the columns in \(J\), so the right-hand side remains \(r\).  For LC5, the rational relative-interior point satisfies \(\mathsf N_Jt_J=r\) and \((t_J)_j\ge1/416\) for every zero-slack variable.

\begin{table}[t]
\caption{\textbf{Certificate and facet data.}  The reduced and full equality systems define the same affine constraints and therefore give the same HIC support values.  For each witness, the listed value of \(\|\eta\|_1\) is the minimum among dual-optimal certificates for the reduced equality representation.  The nonzero-entry and slack counts refer to the selected minimum-norm certificates.  The projected dimensions refer to the no-\(BC\) correlator space.  All ranks are evaluated over \(\mathbb Q\), and the relative-interior points are verified exactly after rational reconstruction.}
\label{tab:supp-diagnostics}
\begin{ruledtabular}
\begin{tabular}{lcc}
item & LC4 & LC5\\
\hline
hidden-influence variables & $256$ & $1024$\\
reduced equality rows & $132$ & $776$\\
full equality rows & $260$ & $1288$\\
rank of the reduced coefficient matrix & $64$ & $334$\\
rank of the full coefficient matrix & $64$ & $334$\\
rank of the reduced augmented system & $64$ & $334$\\
rank of the full augmented system & $64$ & $334$\\
no-\(BC\) correlator coordinates & $44$ & $134$\\
HIC support value & $6$ & $10$\\
minimum \(\|\eta\|_1\) & \(70\) & \(362\)\\
nonzero entries of \(\eta\) & $22$ & $98$\\
zero-slack variables & $176$ & $888$\\
positive-slack variables & $80$ & $136$\\
projected affine dimension & $44$ & $134$\\
facet dimension & $43$ & $133$\\
\end{tabular}
\end{ruledtabular}
\end{table}

\section{Facet verification}
\label{sec:supp-facet}

\subsection{Projected dimension}
Let ${\mathcal P}=\mathsf M\{t:\mathsf N t=r,t\ge0\}$ be the projected no-\(BC\) hidden-influence polytope.  The equality system admits a strictly positive feasible point: for every early setting, assign weight $1/64$ to each LC4 early-outcome/response-function tuple and $1/128$ to each LC5 tuple.  Consequently nonnegativity does not further reduce the affine hull defined by $\mathsf N t=r$.  Its affine dimension can be computed as
\begin{equation}
\dim({\mathcal P})=
\operatorname{rank}
\begin{pmatrix}
\mathsf N\\
\mathsf M
\end{pmatrix}
-\operatorname{rank}(\mathsf N).
\label{eq:supp-dim}
\end{equation}
Exact rational arithmetic gives \(\dim({\mathcal P})=44\) for LC4.  For LC5, \(\operatorname{rank}(\mathsf N)=334\) and \(\operatorname{rank}(\mathsf N;\mathsf M)=468\), giving \(\dim({\mathcal P})=134\). Thus, neither projection satisfies any additional affine equality in the chosen no-\(BC\) correlator coordinates.

\subsection{Dual slack and the exposed face}
\begin{figure}[t]
\centering
\begin{tabular}{@{}c@{\hspace{0.025\textwidth}}c@{}}
\panelinclude[width=0.47\textwidth]{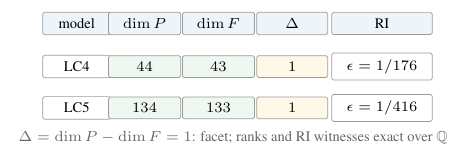}{a} &
\panelinclude[width=0.47\textwidth]{fig_results_comparison.pdf}{b}
\end{tabular}
\caption{\textbf{Facet dimensions and visibility thresholds.}
\panel{a} The exposed faces have dimensions \(43\) and \(133\), one below the corresponding projected dimensions \(44\) and \(134\).
\panel{b} The LC5 visibility threshold is lower than the LC4 threshold in the ideal white-noise model, while LC4 remains the shorter, lower-weight witness.}
\label{fig:supp-summary}
\end{figure}

For either witness, recall the dual slack
{
\begin{equation}
\delta=\mathsf N^T\eta-\alpha.
\label{eq:supp-slack}
\end{equation}
}
Let \(F\) be the face of \({\mathcal P}\) exposed by the witness.  Every feasible weight vector \(t\) that projects to a point in \(F\) satisfies
{
\begin{equation}
\delta^Tt=\mathcal B-\alpha^Tt=0.
\end{equation}
}
Since \(\delta,t\ge0\), \(t_j=0\) whenever \(\delta_j>0\).  Its zero-slack index set is
{
\begin{equation}
J=\{j:\delta_j=0\}.
\end{equation}
}
Restrict \(\mathsf N\) and \(\mathsf M\) to the columns in \(J\).  Every such vector \(t\) is supported on \(J\), so
{
\begin{equation}
\dim(F)\le d_J:=
\operatorname{rank}
\begin{pmatrix}
\mathsf N_J\\
\mathsf M_J
\end{pmatrix}
-\operatorname{rank}(\mathsf N_J).
\label{eq:supp-face-dim}
\end{equation}
}
Exact rational elimination gives \(d_J=43\) for LC4 and \(d_J=133\) for LC5.  The exact relative-interior certificates below show that both upper bounds are attained.
The dimension and relative-interior checks are summarized in Fig.~\ref{fig:supp-summary}(a).

\subsection{Relative-interior feasibility and the facet condition}
Attainment of the support bound does not by itself imply that the exposed face is a facet.  The upper bound \(d_J\) is attained if \(\mathsf N_Jt_J=r\) has a componentwise positive solution.  We therefore solve
{
\begin{equation}
\max \epsilon
\quad\text{subject to}\quad
\mathsf N_Jt_J=r,\quad (t_J)_j\ge\epsilon\ \text{for all }j\in J .
\label{eq:supp-relative-interior}
\end{equation}
}
The certified optima are
{
\begin{equation}
\epsilon_4=\frac{1}{176},
\qquad
\epsilon_5=\frac{1}{416}.
\end{equation}
}
A solution with \(\epsilon>0\) shows that nonnegativity does not reduce the affine hull of \(\{t_J\ge0:\mathsf N_Jt_J=r\}\).  The image of this affine hull has dimension \(d_J\), and \(\delta_J=0\) places the entire restricted image in \(F\).  Hence \(\dim(F)=d_J\).  For LC5, the rational feasible point can be written with common denominator \(3744\) and satisfies the equalities exactly.
The values \(1/176\) and \(1/416\) are lower bounds on the weights \(t_j\) in these feasible decompositions; they are not observed event probabilities.

\begin{proposition}[Facet property]
\label{prop:supp-tightness}
The LC4 and LC5 inequalities define facets of their projected no-\(BC\) hidden-influence polytopes.
\end{proposition}

\begin{proof}
For LC4, \(\dim {\mathcal P}=44\), \(d_J=43\), and Eq.~(\ref{eq:supp-relative-interior}) has a feasible solution with \(\epsilon=1/176>0\).  Hence \(\dim F=43=\dim {\mathcal P}-1\).  For LC5, the corresponding values are \((\dim {\mathcal P},d_J,\epsilon)=(134,133,1/416)\), which gives the same conclusion.  Both inequalities are therefore facet-defining~\cite{Ziegler1995}.
\end{proof}

\section{LC4 stabilizer derivation}
\label{sec:supp-lc4}

\subsection{State, generators and measurements}
The four-qubit linear cluster state is
\begin{equation}
\ket{LC_4}=CZ_{AB}CZ_{BC}CZ_{CD}\ket{+}^{\otimes4}.
\end{equation}
Its stabilizer generators are
\begin{equation}
K_A=X_AZ_B,\quad
K_B=Z_AX_BZ_C,\quad
K_C=Z_BX_CZ_D,\quad
K_D=Z_CX_D .
\label{eq:supp-lc4-generators}
\end{equation}
The measurement choices are
\begin{equation}
A_0=X,\quad A_1=Z,\quad
B_0=\frac{Z+X}{\sqrt2},\quad B_1=\frac{Z-X}{\sqrt2},
\end{equation}
\begin{equation}
C_0=Z,\quad C_1=X,\quad D_0=X,\quad D_1=Z .
\end{equation}
Only $B$ is measured in a rotated basis; the $45^\circ$ rotation is
the source of the $\sqrt2$ factors below.  The rotated pair obeys
\begin{equation}
B_0+B_1=\sqrt2\,Z_B,\qquad
B_0-B_1=\sqrt2\,X_B ,
\label{eq:supp-B-rotation}
\end{equation}
so single $Z_B$ or $X_B$ Paulis are accessible through sums and
differences of the two $B$ settings, while $Y_B$ is not.  Every other
party measures pure $X$ or $Z$, so all accessible observables lie in
the $X$--$Z$ plane.

\begin{figure}[t]
\centering
\begin{tabular}{@{}c@{\hspace{0.03\textwidth}}c@{}}
\panelinclude[width=0.42\textwidth]{fig_lc4_graph.pdf}{a} &
\panelinclude[width=0.48\textwidth]{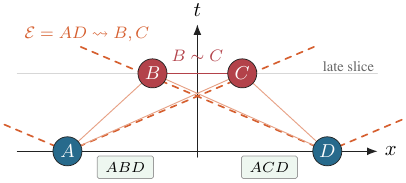}{b}
\end{tabular}
\caption{\textbf{LC4 state and timing.}
\panel{a} Four-qubit linear cluster graph used by the LC4 witness.  The prepared graph contains the edge \(BC\), but the witness contains no joint \(BC\) correlator.
\panel{b} Bancal \textit{et al.}'s timing geometry with early side $\mathcal E=AD$ and mutually blind late pair $B,C$.  The witness is evaluated only from the \(ABD\) and \(ACD\) marginals.}
\label{fig:supp-lc4}
\end{figure}

The LC4 graph and timing geometry are shown in Fig.~\ref{fig:supp-lc4}.

\subsection{Exhaustive enumeration of stabilizer products}
\label{sec:supp-lc4-steps}

The stabilizer group of \(\ket{LC_4}\) contains \(2^4=16\) elements
\(K_S=\prod_{i\in S}K_i\), with \(S\subseteq\{A,B,C,D\}\).
Table~\ref{tab:supp-lc4-all-products} lists the \(15\) nontrivial
products as signed Pauli strings; \(K_S=\pm \mathsf P\) implies
\(\avg{\mathsf P}=\pm1\) on the state.  We retain a product only if it satisfies
both conditions:
\begin{enumerate}[itemsep=2pt,label=(\alph*)]
\item \emph{no-\(BC\) support}: a product whose Pauli string acts
nontrivially on both \(B\) and \(C\) violates the required support
condition;
\item \emph{measurability}: by the discussion below
Eq.~(\ref{eq:supp-B-rotation}), all accessible observables lie in the
\(X\)--\(Z\) plane, so any product containing a \(Y\) Pauli
is inaccessible with the fixed measurements.
\end{enumerate}

\begin{table}[t]
\caption{\textbf{All $15$ nontrivial LC4 stabilizer products.}
Products are reduced using the standard Pauli multiplication rules; repeated \(Z_B\) or \(Z_C\) factors cancel.  The sign convention is
$K_S=\pm \mathsf P$ with $\avg{\mathsf P}=\pm1$ on $\ket{LC_4}$.  ``contains $BC$''
marks failure of the no-\(BC\) support condition; ``needs $Y$'' marks a
no-\(BC\) product that is inaccessible to the fixed $X$--$Z$-plane
settings.  The four retained products generate all six terms of the
witness in Eq.~(\ref{eq:supp-S4}).}
\label{tab:supp-lc4-all-products}
\begin{ruledtabular}
\begin{tabular}{llll}
product & signed Pauli string & status & measurement identity\\
\hline
$K_A$ & $X_AZ_B$ & source of the \(Z_B\) terms in \(R_4\) & $\avg{A_0B_0}+\avg{A_0B_1}=\sqrt2$\\
$K_B$ & $Z_AX_BZ_C$ & contains $BC$ & ---\\
$K_C$ & $Z_BX_CZ_D$ & contains $BC$ & ---\\
$K_D$ & $Z_CX_D$ & term in \(L_4\) & $\avg{C_0D_0}=1$\\
$K_AK_B$ & $Y_AY_BZ_C$ & contains $BC$ & ---\\
$K_AK_C$ & $X_AX_CZ_D$ & term in \(L_4\) & $\avg{A_0C_1D_1}=1$\\
$K_AK_D$ & $X_AZ_BZ_CX_D$ & contains $BC$ & ---\\
$K_BK_C$ & $Z_AY_BY_CZ_D$ & contains $BC$ & ---\\
$K_BK_D$ & $Z_AX_BX_D$ & source of the \(X_B\) terms in \(R_4\) & $\avg{A_1B_0D_0}-\avg{A_1B_1D_0}=\sqrt2$\\
$K_CK_D$ & $Z_BY_CY_D$ & contains $BC$ & ---\\
$K_AK_BK_C$ & $-\,Y_AX_BY_CZ_D$ & contains $BC$ & ---\\
$K_AK_BK_D$ & $Y_AY_BX_D$ & needs $Y$ & ---\\
$K_AK_CK_D$ & $X_AY_CY_D$ & needs $Y$ & ---\\
$K_BK_CK_D$ & $-\,Z_AY_BX_CY_D$ & contains $BC$ & ---\\
$K_AK_BK_CK_D$ & $Y_AX_BX_CY_D$ & contains $BC$ & ---\\
\end{tabular}
\end{ruledtabular}
\end{table}

Among the \(15\) nontrivial products, \(9\) contain both blind parties, \(2\) are no-\(BC\) but require \(Y\) measurements, and \(4\) satisfy both conditions and are retained.  The repeated \(Z_B\) factors in \(K_AK_C\) and the repeated \(Z_C\) factors in \(K_BK_D\) cancel, eliminating \(B\) and \(C\) from the respective Pauli strings.

\subsection{Retained stabilizers and correlator identities}
The four retained products consist of two \(B\)-free stabilizer
terms and two products containing a single \(B\) Pauli.

\emph{\(B\)-free stabilizer terms.}  The two \(B\)-free
products yield deterministic correlators with expectation \(+1\):
\begin{equation}
\avg{C_0D_0}=\avg{Z_CX_D}=\avg{K_D}=1 ,
\end{equation}
and
\begin{equation}
\avg{A_0C_1D_1}=\avg{X_AX_CZ_D}=\avg{K_AK_C}=1 .
\end{equation}

\emph{Terms containing \(B\).}  The two products containing a single $B$
Pauli are converted through Eq.~(\ref{eq:supp-B-rotation}).  The
generator $K_A=X_AZ_B$ gives
\begin{equation}
\avg{A_0B_0}+\avg{A_0B_1}
=\sqrt2\,\avg{X_AZ_B}
=\sqrt2 ,
\end{equation}
and the product $K_BK_D=Z_AX_BX_D$ gives
\begin{equation}
\avg{A_1B_0D_0}-\avg{A_1B_1D_0}
=\sqrt2\,\avg{Z_AX_BX_D}
=\sqrt2 .
\end{equation}
In each rotated-basis expansion, the other Pauli string, obtained by interchanging \(X_B\) and \(Z_B\), has zero expectation because neither the string nor its negative belongs to the stabilizer group of \(\ket{LC_4}\).  The resulting correlators are therefore \(\pm1/\sqrt2\): both terms in the first identity equal \(1/\sqrt2\), while \(\avg{A_1B_0D_0}=1/\sqrt2\) and \(\avg{A_1B_1D_0}=-1/\sqrt2\).  Together, these identities generate all six terms of \(S_4\).

\begin{table}[t]
\caption{\textbf{LC4 correlators.}  The contribution is the coefficient in Eq.~(\ref{eq:supp-S4}) multiplied by the cluster-state expectation value.}
\label{tab:supp-lc4-terms}
\begin{ruledtabular}
\begin{tabular}{lcc}
term & expectation & contribution\\
\hline
$\avg{A_0B_0}$ & $1/\sqrt2$ & $1/\sqrt2$\\
$\avg{A_0B_1}$ & $1/\sqrt2$ & $1/\sqrt2$\\
$\avg{A_1B_0D_0}$ & $1/\sqrt2$ & $1/\sqrt2$\\
$\avg{A_1B_1D_0}$ & $-1/\sqrt2$ & $1/\sqrt2$\\
$\avg{C_0D_0}$ & $1$ & $2$\\
$\avg{A_0C_1D_1}$ & $1$ & $2$\\
\end{tabular}
\end{ruledtabular}
\end{table}

\subsection{Coefficient selection and witness}
The four identities form the blocks \(R_4\) and \(L_4\), with cluster-state point \((R_4^Q,L_4^Q)=(2\sqrt2,2)\).  Section~\ref{sec:supp-construction} shows that the ratio of the cluster-state value to the HIC support is uniquely maximized by \((p,\ell)\propto(1,2)\).  Hence each correlator in the CHSH block has weight \(1\), and each term in the stabilizer block has weight \(2\).

Combining the four stabilizer identities with the weights $(1,2)$
gives the LC4 expression
\begin{equation}
\begin{aligned}
S_4={}&\avg{A_0B_0}+\avg{A_0B_1}
+\avg{A_1B_0D_0}-\avg{A_1B_1D_0}\\
&+2\avg{C_0D_0}+2\avg{A_0C_1D_1}.
\end{aligned}
\label{eq:supp-S4}
\end{equation}
The first two terms are supported on $AB$, the next two on $ABD$, the
fifth on $CD$, and the last on $ACD$.  The expression is therefore
no-\(BC\).  The integer Farkas certificate of
Sec.~\ref{sec:supp-certificates} proves the hidden-influence bound
$S_4\le6$.  Section~\ref{sec:supp-facet} shows that this bound is
attained and that its exposed face is a facet.

\subsection{Quantum value, bounds, and noise threshold}
Table~\ref{tab:supp-lc4-terms} lists the cluster-state expectation value
of each term, as obtained from the correlator identities above.

Summing Table~\ref{tab:supp-lc4-terms} gives
\begin{equation}
S_4^Q=4+2\sqrt2\approx6.8284271247.
\end{equation}
The ideal white-noise threshold is
\begin{equation}
\nu_4^*=\frac{6}{4+2\sqrt2}\approx0.8786796564.
\end{equation}

At the cluster-state point, the rotated \(B\) block contributes \(2\sqrt2\) and the \(B\)-free stabilizer terms contribute \(4\).  The no-signaling and HIC support values are \(8\) and \(6\), respectively, so the stronger bound depends on conditional \(BC\) locality in addition to no-signaling.
Assuming measurement independence, no postselection, the stated spacetime arrangement, and operational no-signaling, an observed value \(S_4>6\) is incompatible with the HIC constraints.  Both the no-signaling and algebraic maxima equal \(8\).

\begin{figure}[t]
\centering
\begin{tabular}{@{}c@{\hspace{0.03\textwidth}}c@{}}
\panelinclude[width=0.42\textwidth]{fig_lc5_graph.pdf}{a} &
\panelinclude[width=0.48\textwidth]{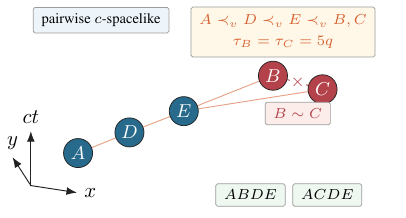}{b}
\end{tabular}
\caption{\textbf{LC5 state and timing.}
\panel{a} Five-qubit linear cluster graph with an additional qubit \(E\) on the early side.
\panel{b} The five-site arrangement in Eq.~(\ref{eq:supp-lc5-spacetime-coordinates}), schematically shown for \({\kappa}=2\) and \({\zeta}=1/20\). The orange arrows show \(A\prec_v D\prec_v E\prec_v B,C\), with \(B\sim C\).  The \(C\)-delayed branch fixes the target \(ABDE\) marginal, while the \(B\)-delayed branch fixes the target \(ACDE\) marginal.}
\label{fig:supp-lc5}
\end{figure}

\section{LC5 stabilizer derivation}
\label{sec:supp-lc5}

\subsection{State, generators and measurements}
The five-qubit linear cluster state is
\begin{equation}
\ket{LC_5}=CZ_{AB}CZ_{BC}CZ_{CD}CZ_{DE}\ket{+}^{\otimes5}.
\end{equation}
Its stabilizer generators are
\begin{equation}
\begin{aligned}
K_A&=X_AZ_B,\qquad
K_B=Z_AX_BZ_C,\qquad
K_C=Z_BX_CZ_D,\\
K_D&=Z_CX_DZ_E,\qquad
K_E=Z_DX_E .
\end{aligned}
\label{eq:supp-lc5-generators}
\end{equation}
The measurements extend those of LC4 by the fifth qubit:
\begin{equation}
A_0=X,\quad A_1=Z,\quad
B_0=\frac{Z+X}{\sqrt2},\quad B_1=\frac{Z-X}{\sqrt2},
\end{equation}
\begin{equation}
C_0=Z,\quad C_1=X,\quad D_0=X,\quad D_1=Z,\quad E_0=Z,\quad E_1=X .
\end{equation}
As in LC4, only \(B\) is measured in a rotated basis.  Equation~(\ref{eq:supp-B-rotation}) gives access to correlators containing \(Z_B\) or \(X_B\), but not \(Y_B\).  Together with the \(X\)- or \(Z\)-measurements at the other sites, this restricts all accessible observables to the \(X\)--\(Z\) plane.
The LC5 graph and timing geometry are shown in Fig.~\ref{fig:supp-lc5}.

\subsection{Exhaustive enumeration of stabilizer products}
\label{sec:supp-lc5-steps}
The stabilizer group of \(\ket{LC_5}\) has \(2^5=32\) elements \(K_S=\prod_{i\in S}K_i\), with \(S\subseteq\{A,B,C,D,E\}\).  Table~\ref{tab:supp-lc5-all-products} lists the \(31\) nontrivial products and identifies those satisfying the no-\(BC\) support and \(X\)--\(Z\)-measurability conditions of Sec.~\ref{sec:supp-lc4-steps}.

\begin{table}[t]
\caption{\textbf{All $31$ nontrivial LC5 stabilizer products.}
Sign convention $K_S=\pm \mathsf P$ with $\avg{\mathsf P}=\pm1$ on $\ket{LC_5}$.
``contains $BC$'' marks failure of the no-\(BC\) support condition;
``needs $Y$'' marks a no-\(BC\) product inaccessible to the fixed
settings.  The seven retained products generate all
stabilizer-derived terms of the witness in Eq.~(\ref{eq:supp-S5});
the four completion terms of Eq.~(\ref{eq:supp-S5}) are not
stabilizer products and enter through the dichotomic completion,
Eq.~(\ref{eq:supp-effective-A-lc5}).}
\label{tab:supp-lc5-all-products}
\begin{ruledtabular}
\begin{tabular}{llll}
product & signed Pauli string & status & measurement identity\\
\hline
$K_A$ & $X_AZ_B$ & source of the \(Z_B\) terms in \(R_5\) & $\avg{A_0B_0}+\avg{A_0B_1}=\sqrt2$\\
$K_B$ & $Z_AX_BZ_C$ & contains $BC$ & ---\\
$K_C$ & $Z_BX_CZ_D$ & contains $BC$ & ---\\
$K_D$ & $Z_CX_DZ_E$ & term in \(L_{5,2}\) & $\avg{C_0D_0E_0}=1$\\
$K_E$ & $Z_DX_E$ & term in \(L_{5,1}\) & $\avg{D_1E_1}=1$\\
$K_AK_B$ & $Y_AY_BZ_C$ & contains $BC$ & ---\\
$K_AK_C$ & $X_AX_CZ_D$ & term in \(L_{5,1}\) & $\avg{A_0C_1D_1}=1$\\
$K_AK_D$ & $X_AZ_BZ_CX_DZ_E$ & contains $BC$ & ---\\
$K_AK_E$ & $X_AZ_BZ_DX_E$ & source of the \(Z_B\) terms in \(R_5\) & $\avg{A_0B_0D_1E_1}+\avg{A_0B_1D_1E_1}=\sqrt2$\\
$K_BK_C$ & $Z_AY_BY_CZ_D$ & contains $BC$ & ---\\
$K_BK_D$ & $Z_AX_BX_DZ_E$ & source of the \(X_B\) terms in \(R_5\) & $\avg{A_1B_0D_0E_0}-\avg{A_1B_1D_0E_0}=\sqrt2$\\
$K_BK_E$ & $Z_AX_BZ_CZ_DX_E$ & contains $BC$ & ---\\
$K_CK_D$ & $Z_BY_CY_DZ_E$ & contains $BC$ & ---\\
$K_CK_E$ & $Z_BX_CX_E$ & contains $BC$ & ---\\
$K_DK_E$ & $Z_CY_DY_E$ & needs $Y$ & ---\\
$K_AK_BK_C$ & $-\,Y_AX_BY_CZ_D$ & contains $BC$ & ---\\
$K_AK_BK_D$ & $Y_AY_BX_DZ_E$ & needs $Y$ & ---\\
$K_AK_BK_E$ & $Y_AY_BZ_CZ_DX_E$ & contains $BC$ & ---\\
$K_AK_CK_D$ & $X_AY_CY_DZ_E$ & needs $Y$ & ---\\
$K_AK_CK_E$ & $X_AX_CX_E$ & term in \(L_{5,1}\) & $\avg{A_0C_1E_1}=1$\\
$K_AK_DK_E$ & $X_AZ_BZ_CY_DY_E$ & contains $BC$ & ---\\
$K_BK_CK_D$ & $-\,Z_AY_BX_CY_DZ_E$ & contains $BC$ & ---\\
$K_BK_CK_E$ & $Z_AY_BY_CX_E$ & contains $BC$ & ---\\
$K_BK_DK_E$ & $Z_AX_BY_DY_E$ & needs $Y$ & ---\\
$K_CK_DK_E$ & $-\,Z_BY_CX_DY_E$ & contains $BC$ & ---\\
$K_AK_BK_CK_D$ & $Y_AX_BX_CY_DZ_E$ & contains $BC$ & ---\\
$K_AK_BK_CK_E$ & $-\,Y_AX_BY_CX_E$ & contains $BC$ & ---\\
$K_AK_BK_DK_E$ & $Y_AY_BY_DY_E$ & needs $Y$ & ---\\
$K_AK_CK_DK_E$ & $-\,X_AY_CX_DY_E$ & needs $Y$ & ---\\
$K_BK_CK_DK_E$ & $Z_AY_BX_CX_DY_E$ & contains $BC$ & ---\\
$K_AK_BK_CK_DK_E$ & $-\,Y_AX_BX_CX_DY_E$ & contains $BC$ & ---\\
\end{tabular}
\end{ruledtabular}
\end{table}

Of the \(31\) nontrivial products, \(18\) contain both blind parties, \(6\) require \(Y\) measurements, and \(7\) satisfy both conditions and are retained.  As in LC4, repeated \(Z_B\) or \(Z_C\) factors cancel, eliminating one of \(B\) and \(C\) from the corresponding Pauli string.  The additional generator \(K_E\) yields the retained products \(K_E\), \(K_AK_E\), and \(K_AK_CK_E\).

\subsection{Retained stabilizers and correlator identities}
The seven retained products consist of four \(B\)-free stabilizer terms and three products containing a single \(B\) Pauli.  The \(B\)-free terms have expectation \(+1\), while Eq.~(\ref{eq:supp-B-rotation}) splits each single-\(B\) Pauli into two correlators.  The four \(B\)-free products give
\begin{align}
\avg{C_0D_0E_0}=\avg{K_D}=1,&\qquad
\avg{D_1E_1}=\avg{K_E}=1,\nonumber \\
\avg{A_0C_1D_1}=\avg{K_AK_C}=1,&\qquad
\avg{A_0C_1E_1}=\avg{K_AK_CK_E}=1,
\end{align}
and the three CHSH-block components give
\begin{align}
K_A=X_AZ_B:\quad
&\avg{A_0B_0}+\avg{A_0B_1}=\sqrt2 ,\\
K_AK_E=X_AZ_BZ_DX_E:\quad
&\avg{A_0B_0D_1E_1}+\avg{A_0B_1D_1E_1}=\sqrt2 ,\\
K_BK_D=Z_AX_BX_DZ_E:\quad
&\avg{A_1B_0D_0E_0}-\avg{A_1B_1D_0E_0}=\sqrt2 .
\end{align}
For each pair, the other Pauli string, obtained by interchanging \(X_B\) and \(Z_B\), has zero expectation because neither the string nor its negative belongs to the stabilizer group of \(\ket{LC_5}\).
The individual correlators therefore take the stated \(\pm1/\sqrt2\) values.
Separately, because \(D_1E_1=Z_DX_E=K_E\), \(\avg{D_1E_1}\) arises from a stabilizer, not from the dichotomic completion.

\subsection{Role of the completion terms}
The four additional terms
\begin{equation}
-\avg{B_0D_1}-\avg{B_1D_1}+\avg{B_0E_1}+\avg{B_1E_1},
\end{equation}
are not stabilizers.  Resolving \(B_0\) and \(B_1\) into their \(Z_B\) and \(X_B\) components gives the four Pauli strings \(Z_BZ_D\), \(X_BZ_D\), \(Z_BX_E\), and \(X_BX_E\).  For each of these four Pauli strings \(\mathsf P\), neither \(\mathsf P\) nor \(-\mathsf P\) belongs to the stabilizer group of \(\ket{LC_5}\), so all four completion terms have zero expectation.  These terms arise from completing the quantity \(A_0(1+D_1E_1)/2\), which vanishes for \(D_1=-E_1\), to the dichotomic variable \(\widetilde A_0\) in Eq.~(\ref{eq:supp-effective-A-lc5}).  Deleting all four leaves the HIC support value equal to \(10\) but yields an inequality that is not facet-defining, as shown below.

Let \(S_5'\) denote Eq.~(\ref{eq:supp-S5}) with these four
terms deleted, and let \(\alpha'\) be its objective vector.
An exact feasible vector \(t'\) with eight nonzero entries satisfies
\(\alpha'^Tt'=10\).  A refined integer dual
\(\widetilde\eta\), with \(175\) nonzero entries and
\(\|\widetilde\eta\|_1=314\), satisfies
{
\begin{equation}
\mathsf N^T\widetilde\eta-\alpha'\ge0,
\qquad
r^T\widetilde\eta=10.
\label{eq:supp-lc5-reduced-dual}
\end{equation}
}
Together, the matching primal and dual values establish that the HIC support value of \(S_5'\) is \(10\).  Define
\(\widetilde\delta=\mathsf N^T\widetilde\eta-\alpha'\) and
\(K=\{j:\widetilde\delta_j=0\}\).  The set \(K\) contains \(744\) indices.  An exact rational vector
\(t^{\mathrm{RI}}=k/416\), supported on \(K\), is strictly positive on every index in \(K\) and satisfies
{
\begin{equation}
\begin{gathered}
\mathsf N_Kt_K^{\mathrm{RI}}=r,
\qquad
(t_K^{\mathrm{RI}})_j\ge\frac1{416}\quad(j\in K),
\qquad
\alpha'^Tt^{\mathrm{RI}}=10,\\
\operatorname{rank}_{\mathbb Q}(\mathsf N_K)=326,
\qquad
\operatorname{rank}_{\mathbb Q}
\begin{pmatrix}
\mathsf N_K\\
\mathsf M_K
\end{pmatrix}
=451.
\end{gathered}
\label{eq:supp-lc5-reduced-ranks}
\end{equation}
}
Let \(F'\) be the face exposed by \(S_5'\).  For any point in \(F'\)
and any feasible weight vector \(t\) that projects to it,
{
\begin{equation*}
0=10-\alpha'^Tt=\widetilde\delta^Tt.
\end{equation*}
}
Since \(\widetilde\delta,t\ge0\), every such vector \(t\) is supported on \(K\).  Conversely, \(\widetilde\delta_K=0\), so every feasible vector supported on \(K\) saturates the inequality.  The componentwise-positive vector \(t^{\mathrm{RI}}\) shows that nonnegativity does not further reduce the affine hull of this subsystem.  Therefore the rank difference gives the exact
exposed-face dimension:
{
\begin{equation}
\dim F'=451-326=125,
\qquad
\operatorname{codim}_{\mathcal P}F'=134-125=9.
\label{eq:supp-lc5-reduced-face}
\end{equation}
}
Thus \(S_5'\) is not facet-defining; a facet would require dimension \(133\).  Since the deleted terms vanish on \(\ket{LC_5}\), \(S_5'\) has the same cluster-state value and ideal white-noise threshold as \(S_5\).

\subsection{Coefficient selection and witness}
Section~\ref{sec:supp-construction} gives the primitive block normal \((1,1,2)\).  It fixes the coefficients multiplying \((R_5,L_{5,1},L_{5,2})\) in the ratio \(1:1:2\), corresponding to the coefficient pattern \(1,2,4\) in the expanded witness.

Combining the blocks with the weights $(1,1,2)$ gives the LC5
expression
\begin{equation}
\begin{aligned}
S_5={}&
\avg{A_0B_0}+\avg{A_0B_1}
-\avg{B_0D_1}-\avg{B_1D_1}
+\avg{B_0E_1}+\avg{B_1E_1}\\
&+2\avg{A_0C_1D_1}
+2\avg{A_0C_1E_1}
-2\avg{D_1E_1}\\
&+\avg{A_0B_0D_1E_1}
+\avg{A_0B_1D_1E_1}\\
&+2\avg{A_1B_0D_0E_0}
-2\avg{A_1B_1D_0E_0}
+4\avg{C_0D_0E_0}.
\end{aligned}
\label{eq:supp-S5}
\end{equation}
Every term belongs to an \(ABDE\) or \(ACDE\) marginal, or to one of their submarginals, so none contains both \(B\) and \(C\).  As
for LC4, the integer Farkas certificate of
Sec.~\ref{sec:supp-certificates} proves the hidden-influence bound
$S_5\le10$.  Section~\ref{sec:supp-facet} shows that this bound is
attained and that its exposed face is a facet.

\subsection{Quantum value, noise threshold, and comparison with LC4}
Table~\ref{tab:supp-lc5-terms} lists the cluster-state expectations and their contributions to \(S_5\).  The four completion terms vanish, whereas the stabilizer-derived terms sum to \(6+4\sqrt2\).

\begin{table}[t]
\caption{\textbf{LC5 correlators.}  The contribution column includes the coefficient in Eq.~(\ref{eq:supp-S5}). The four zero-expectation completion terms are the two-body terms in Eq.~(\ref{eq:supp-R5-expanded}) generated by the dichotomic completion in Eq.~(\ref{eq:supp-effective-A-lc5}); deleting all four yields the non-facet-defining expression \(S_5'\).}
\label{tab:supp-lc5-terms}
\begin{ruledtabular}
\begin{tabular}{lcc}
term(s) & expectation value(s) & contribution\\
\hline
$\avg{A_0B_0}+\avg{A_0B_1}$ & $1/\sqrt2,1/\sqrt2$ & $\sqrt2$\\
$-\avg{B_0D_1}-\avg{B_1D_1}$ & $0,0$ & $0$\\
$\avg{B_0E_1}+\avg{B_1E_1}$ & $0,0$ & $0$\\
$2\avg{A_0C_1D_1}$ & $1$ & $2$\\
$2\avg{A_0C_1E_1}$ & $1$ & $2$\\
$-2\avg{D_1E_1}$ & $1$ & $-2$\\
$\avg{A_0B_0D_1E_1}+\avg{A_0B_1D_1E_1}$ & $1/\sqrt2,1/\sqrt2$ & $\sqrt2$\\
$2\avg{A_1B_0D_0E_0}-2\avg{A_1B_1D_0E_0}$ & $1/\sqrt2,-1/\sqrt2$ & $2\sqrt2$\\
$4\avg{C_0D_0E_0}$ & $1$ & $4$\\
\end{tabular}
\end{ruledtabular}
\end{table}

Summing the contributions gives
\begin{equation}
S_5^Q=6+4\sqrt2\approx11.6568542495,
\end{equation}
and the ideal white-noise threshold is
\begin{equation}
\nu_5^*=\frac{10}{6+4\sqrt2}\approx0.8578643763.
\end{equation}
Compared with LC4, LC5 enlarges the early side from \(AD\) to \(ADE\) and the projected no-\(BC\) space from \(44\) to \(134\) dimensions.  Its distributed CHSH block is supported within \(ABDE\), whereas its
\(C\)-side lock terms are supported within \(ACDE\) and its submarginals.  LC4 uses six two- and three-body correlators and no four-body terms; LC5 has the lower ideal white-noise threshold but includes four-body \(ABDE\) correlators.
The ideal white-noise thresholds are compared in Fig.~\ref{fig:supp-summary}(b).

\section{Noise, sampling, and experimental use}
\label{sec:supp-robustness}

\subsection{White noise}
Consider
\begin{equation}
P_{\nu}=\nu P_Q+(1-\nu)P_{\mathrm{white}},
\end{equation}
where \(P_Q\) denotes the target quantum distribution and \(P_{\mathrm{white}}\) the uniform white-noise distribution.  Every nonempty correlator in $S_4$ and $S_5$ is multiplied by $\nu$.  Hence the threshold is exactly
\begin{equation}
\nu^*=\frac{\mathcal B}{S^Q}.
\end{equation}
The values are
\begin{equation}
\nu_4^*\approx0.8786796564,\qquad
\nu_5^*\approx0.8578643763.
\end{equation}
Equivalently, the ideal white-noise tolerances are about $12.13\%$ and $14.21\%$.

\subsection{Readout-error sensitivity}
Suppose each raw binary output is flipped independently with symmetric probability \(p_{\rm flip}\).  The expected observed \(k\)-body correlator is then
\begin{equation}
\avg{O_1\cdots O_k}_{\mathrm{obs}}
=(1-2p_{\rm flip})^k\avg{O_1\cdots O_k}_{\mathrm{ideal}}.
\label{eq:supp-readout}
\end{equation}
Equation~(\ref{eq:supp-readout}) is used only to predict the observed witness value; no inverse readout correction enters the certification.
LC4 contains two- and three-body correlators.  LC5 contains two-, three-, and four-body correlators.  Under this independent symmetric-flip model, the LC4 observed value is
\begin{equation}
\begin{aligned}
S_{4,\mathrm{obs}}{=}{}&
(1-2p_{\rm flip})^2\left(\avg{A_0B_0}_{\mathrm{ideal}}+\avg{A_0B_1}_{\mathrm{ideal}}
+2\avg{C_0D_0}_{\mathrm{ideal}}\right)\\
&+(1-2p_{\rm flip})^3\left(\avg{A_1B_0D_0}_{\mathrm{ideal}}
-\avg{A_1B_1D_0}_{\mathrm{ideal}}
+2\avg{A_0C_1D_1}_{\mathrm{ideal}}\right).
\end{aligned}
\end{equation}
LC5 has additional four-body terms with attenuation $(1-2p_{\rm flip})^4$.  The lower LC5 white-noise threshold therefore need not imply greater robustness to readout errors.

\subsection{Shot-noise propagation}
For correlator \(j\), let \(Y_{j,1},\ldots,Y_{j,N_j}\in\{\pm1\}\) be independent and identically distributed outcomes with mean \(m_j\), and set \(\widehat m_j=N_j^{-1}\sum_nY_{j,n}\).  Then
{
\begin{equation}
\operatorname{Var}(\widehat m_j)
=\frac{1-m_j^2}{N_j}
\le\frac1{N_j}.
\label{eq:supp-correlator-var}
\end{equation}
}
For a witness \(S=\sum_j\omega_jm_j\) whose terms are estimated from statistically independent shot sets,
{
\begin{equation}
\operatorname{Var}(\widehat S)
=\sum_j\omega_j^2\frac{1-m_j^2}{N_j}
\le\sum_j\frac{\omega_j^2}{N_j}.
\label{eq:supp-shot-var}
\end{equation}
}
If several correlators are estimated from the same setting records, their estimators may be correlated, in which case
\begin{equation}
\operatorname{Var}(\widehat{S})
=\sum_{j,k}\omega_j\omega_k\operatorname{Cov}(\widehat m_j,\widehat m_k).
\label{eq:supp-shot-covariance}
\end{equation}
Equation~(\ref{eq:supp-shot-var}) applies only to independent estimators; otherwise Eq.~(\ref{eq:supp-shot-covariance}) or a valid upper bound must be used.

For independent term estimators, set \(\sigma_j^2=1-m_j^2\).  Restrict the optimization to terms with \(\sigma_j>0\), and let \(N_{\rm tot}\) denote the shot budget assigned to these terms.  Treating their shot counts as positive continuous variables, minimizing Eq.~(\ref{eq:supp-shot-var}) gives
{
\begin{equation}
N_j=N_{\rm tot}
\frac{|\omega_j|\sigma_j}
{\sum_{k:\,\sigma_k>0}|\omega_k|\sigma_k},
\qquad \sigma_j>0.
\label{eq:supp-optimal-shot-allocation}
\end{equation}
}
The continuous allocation in Eq.~(\ref{eq:supp-optimal-shot-allocation}) must be rounded to integer counts while preserving the total budget and any minimum allocation per term.  Terms with \(\sigma_j=0\) do not enter the variance minimization and must be assigned separately if they are to be measured.  If the variances are unknown, the conservative choice \(\sigma_j=1\) gives \(N_j\propto|\omega_j|\).

\bibliography{references}